\documentclass[prb,showpacs,twocolumn]{revtex4-1}

\usepackage{graphicx,amssymb,amsmath,color}

\begin{document}
\title{Edge effects in graphene nanostructures: \\ I. From multiple reflection expansion to density of states}

\author{J\"urgen Wurm}
\author{Klaus Richter}
\affiliation{Institut f\"ur Theoretische Physik, Universit\"at Regensburg, D-93040 Regensburg, Germany}

\author{\.{I}nan\c{c} Adagideli}
\affiliation{Faculty of Engineering and Natural Sciences, Sabanc\i~ University, Orhanl\i~ - Tuzla, 34956, Turkey}

\date{\today}

\newcommand{\bsy}[1]{\boldsymbol{#1}}
\newcommand{\imi}{i}
\newcommand{\bra}[1]{\langle #1 |}
\newcommand{\ket}[1]{|#1\rangle}
\newcommand{\braket}[2]{\langle #1 |#2 \rangle}
\newcommand{\mypara}[1]{\paragraph{#1}~\\}
\newcommand{\sgn}{\text{sgn}}
\newcommand{\Gsc}{G^{\text{sc}}}
\newcommand{\sigmat}{\sigma_{\text{t}}}
\newcommand{\sigman}{\sigma_{\text{n}}}
\newcommand{\sigmaa}{\sigma_{\bsy{a}}}
\newcommand{\sigmab}{\sigma_{\bsy{b}}}
\newcommand{\tr}{\text{Tr}}
\newcommand{\taud}{\tau_{\text{d}}}
\newcommand{\taudexp}{{\displaystyle\tau_{\text{d}}}}
\newcommand{\thetal}{\theta_{\text{loop}}}
\newcommand{\bsig}{\bsy{\sigma}}
\newcommand{\dagg}{^{\dagger}}
\newcommand{\cdotsh}{\!\cdot\!}
\newcommand{\vF}{v_{\scriptscriptstyle F}}
\newcommand{\kE}{k_{\scriptscriptstyle E}}
\newcommand{\kmax}{k_\text{max}}
\newcommand{\sigmav}[1]{\sigma_{#1}}
\newcommand{\sDelta}{{\scriptstyle \Delta}}
\newcommand{\OpI}{\hat{\mathcal{I}}}
\newcommand{\I}{{\mathcal{I}}}
\newcommand{\OpIs}{\hat{\mathcal{I}}_{\text{s}}}
\newcommand{\Is}{{\mathcal{I}}_{\text{s}}}
\newcommand{\OpIl}{\hat{\mathcal{I}}_{\text{l}}}
\newcommand{\Il}{{\mathcal{I}}_{\text{l}}}
\newcommand{\intl}{\int\limits}
\newcommand{\unit}[1]{\,\text{#1}}
\newcommand{\rhoosc}{\rho_{\text{osc}}}
\newcommand{\rhobar}{\bar{\rho}}
\newcommand{\Ktilde}{\tilde{K}}
\newcommand{\Ktildes}{\tilde{K}^{\text{s}}}
\newcommand{\Ktildesd}{\tilde{K}^{\text{s}\dagger}}
\newcommand{\Ktildev}{\tilde{K}^{\text{v}}}
\newcommand{\Ktildevd}{\tilde{K}^{\text{v}\dagger}}
\newcommand{\disptauac}{{\displaystyle \tau_{\text{ac}}}}
\newcommand{\disptaud}{{\displaystyle \tau_{\text{d}}}}
\newcommand{\disptauB}{{\displaystyle \tau_{\text{B}}}}
\newcommand{\Wac}{W_{\text{ac}}}
\newcommand{\Tmin}{T_{\text{min}}(\varepsilon)}

\begin{abstract}

We study the influence of different edge types on the electronic density of states of graphene nanostructures. To this end we develop an exact expansion for the single particle Green's function of ballistic graphene structures in terms of multiple reflections from the system boundary, that allows for a natural treatment of edge effects. We first apply this formalism to calculate the average density of states of graphene billiards. While the leading term in the corresponding Weyl expansion is proportional to the billiard area, we find that the contribution that usually scales with the total length of the system boundary differs significantly from what one finds in semiconductor-based, Schr\"odinger type billiards: The latter term vanishes for armchair and infinite mass edges and is proportional to the zigzag edge length, highlighting the prominent role of zigzag edges in graphene. We then compute analytical expressions for the density of states oscillations and energy levels within a trajectory based semiclassical approach. We derive a Dirac version of Gutzwiller's trace formula for classically chaotic graphene billiards and further obtain semiclassical trace formulae for the density of states oscillations in regular graphene cavities. We find that edge dependent interference of pseudospins in graphene crucially affects the quantum spectrum.

\end{abstract}


\maketitle
\section{Introduction}

\subsection{Graphene-based nanostructures}
Triggered by the experimental discovery of massless Dirac quasiparticles\cite{Novoselov2004, Novoselov2005}, graphene has become one of the most intensively studied materials of the last decade (for reviews on physical properties see Refs. \onlinecite{Geim2007, Avouris2007, Beenakker2008, Castro2009, Abergel2010}).

Subsequently, graphene-based nanostructures have been the focus of an immense experimental activity, including graphene nanoribbons\cite{Han2007, Li2008, Tapaszto2008, Gallagher2010}, 
quantum dots \cite{Ponomarenko2008, Guttinger2008, Guttinger2010}, 
Aharonov-Bohm rings\cite{Russo2008, Huefner2010} and antidot arrays\cite{Eroms2009, Bai2010}, raising the issue of confining massless Dirac electrons. 
On the theoretical side, several studies have also focused on graphene nanostructures: Graphene nanoribbons have been studied first using a lattice model\cite{Fujita1996, Nakada1996}. The wavefunctions and energy spectra of graphene nanoribbons have been derived by Brey and Fertig\cite{Brey2006} for armchair and zigzag type edges, and by Tworzyd\l o and coworkers\cite{Tworzydlo2006} for the case of infinite mass edges. The spectral and transport properties of Dirac electrons confined in graphene quantum dots have been investigated analytically\cite{Silvestrov2007, Trauzettel2007, Recher2009} and by numerical means\cite{Bardarson2009, Libisch2009,  Wurm2009, Wimmer2010}. Also energy spectrum and conductance of Aharonov-Bohm rings have been the focus of several publications\cite{Recher2007, Wurm2010, Schelter2010} as well as superlattice effects in graphene antidot lattices\cite{Pedersen2008, Vanevic2009} and the density of states of nanoribbon-superconductor junctions\cite{Herrera2010}.

One upshot of these studies is the understanding that the confinement of charge carriers in graphene affects the coherent electron and hole dynamics considerably. In conventional two-dimensional electron systems (2DES) such as low-dimensional semiconductor structures, the charge carriers can be confined, e.g. by the application of top or side gate voltages, and the quasiparticle transport does not depend on the minute details of the resulting effective potential.
In contrast, in graphene, electrostatic potentials do not necessarily confine charge carriers as the Dirac spectrum does not have a gap\cite{Beenakker2008}.
Thus the confined electrons or holes in graphene nanostructures or flakes are expected to scatter from the very ends of the terminated graphene lattice, and the internal degrees of freedom (such as spin or pseudospin) of the quasiparticles before and after the scattering are considerably affected by the atomic level details of the edges. This mixing of internal (pseudo)spin with orbital degrees of freedom of charge carriers at the boundary leads to richer boundary conditions than for the conventional 2DES\cite{McCann2004, Akhmerov2007, Akhmerov2008}. These boundary conditions in turn affect the spectral and transport properties. However, experimental control and manipulation of edges at an atomistic level is far from being achieved.
Thus a full theoretical description is desirable. However, the edge disorder differs from usual (weak) bulk disorder in that weak coupling perturbation theories cannot treat edges. Therefore this paper is dedicated to develop a formalism that includes the effects of edges non-perturbatively, and to subsequently apply this formalism to study edge effects on the spectral density of states of graphene nanostructures.
\subsection{Scope of this work}

Cutting a finite piece of graphene out of the bulk will generally lead
to disordered boundaries with local properties depending on the respective
orientation of an edge segment with respect to the crystallographic axes.
The accurate calculation of the eigenenergies these finite graphene
systems usually requires numerical quantum mechanical approaches. However, it appears difficult to systematically
resolve edge phenomena from other quantum effects or to unravel generic features 
of graphene nanostructures using numerical simulations. Here we follow a complementary strategy:
We adapt the multiple reflection expansion\cite{Balian1970, Adagideli2002}, { i.\,e.} a representation of the 
Green's function in terms of the number of reflections from the system boundaries, 
to the case of graphene. We thus incorporate edge effects (due to armchair, zigzag and
infinite mass type and combinations of such edge segments) in a direct and
transparent way. We next derive a semiclassical approximation for the Green's function, assuming the Fermi wavelength is much smaller that the typical system size $L$, i.\,e. 
$L \gg 1/\kE$. On the other hand, the Dirac equation that we use is valid for Fermi wavelengths that are large compared to the lattice constant $a\approx 2.46$\,\AA\,, i.\,e. if $1/\kE \gg a$. For mesoscopic systems with $L\gg a$, the semiclassical approximation can thus be well fulfilled in the linear dispersion regime, in which quasiparticle dynamics is governed by the effective Dirac equation. The resulting Green's function then can be used to calculate the density of states (DOS) or the conductance, and their correlators.

In this work we consider the density of states. We focus on gross structures and spectral densities arising from moderate smearing
of the level density and on the calculation of DOS oscillations and individual levels separately. To this end we decompose the DOS into
an average part and the remaining oscillatory contribution. The average spectral density, approximated by the so-called Weyl expansion \cite{Weyl1911, Balian1970, Gutzwiller1990} valid in the semiclassical limit, is a fundamental quantity
of a cavity. It incorporates various geometrical and quantum features, including edge effects. For billiards with spin-orbit interaction, the smooth part of the engery spectrum has been studied in Ref.\,\onlinecite{Cserti2004}.
The oscillatory part of the DOS is computed by invoking 
a semiclassical approximation, leading to so-called semiclassical trace formulae,
{i.\,e.} sums over coherent amplitudes associated to classical periodic orbits.
For graphene cavities with shapes giving rise to regular or chaotic classical dynamics we
derive trace formulae analogous to those known 
(Berry-Tabor \cite{Berry1976/77} and Gutzwiller \cite{Gutzwiller1990} formula, respectively)
for the corresponding Schr\"odinger billiards, i.\,e. billiard systems based on the 
Schr\"odinger equation with Dirichlet boundary conditions. For two representative regular shapes, 
we compute the DOS oscillations and the semiclassical energy levels explicitly. 
The effects of both, the underlying effective Dirac equation (for graphene close to the 
Dirac point) and reflections at different kinds of edges, is incorporated by a pseudospin propagator associated with each orbit, multiplying the usual semiclassical amplitude. Semiclassical trace
formulae involving the electron spin dynamics have been earlier considered for the massive
Dirac equation by Bolte and Keppeler \cite{Bolte1999} and for bulk graphene by Carmier and 
Ullmo \cite{Carmier2008}. Related trace formulae appear also in trajectory-based treatments
of electronic systems with spin-orbit interaction \cite{Pletyukhov2002, Chang2004, Zaitsev2005, Adagideli2010}.
We note that semiclassical methods have also been used to study graphene in magnetic fields\cite{Kormanyos2008, Rakyta2010, Carmier2010}.

Following the concepts outlined above we address edge effects on the 
electronic spectra of closed graphene cavities and quantum transport through 
open graphene systems in two consecutive papers. In the present paper we first
derive the single-particle Green's function and its semiclassical approximation
for graphene cavities and calculate the density of states. 
In subsequent work\cite{partII} we will consider quantities based on products of 
single-particle Green's functions. They include the transport quantities such as the conductance as well as the spectral 
two-point correlator and its dual the spectral form factor, as a tool to study 
spectral statistics. The semiclassical treatment of observables
based on products of Green's functions requires additional techniques which builds
the conceptual basis of the second paper\cite{partII}.

The present paper is organized as follows: After introducing below the effective
Hamiltonian and (matrix) boundary conditions for the different edge types, we derive in Sec. \ref{sec:mre} the multiple reflection expansion (MRE) for the Green's function of a ballistic graphene structure. With this expansion as a starting point we then compute in Sec. \ref{sec:Weyl} the first two terms in the Weyl expansion for the smooth part of the DOS of graphene billiards, particularly focusing on contributions from the boundary. We compare our analytical theory with numerical quantum simulations for various graphene billiards with different edge structures.
In Sec. \ref{sec:OscillatingDOS} we turn to the oscillatory part of the DOS . To this end we first obtain a general semiclassical approximation to the MRE for the graphene Green's function in terms of sums over classical trajectories in \ref{ssec:Semiclassics}. Subsequently we focus on the DOS oscillations in graphene billiards with regular classical dynamics in \ref{ssec:regular}. We give semiclassical trace formulae for two exemplary geometries, namely disks and rectangles, and discuss the effects of the graphene edges. Finally we extend Gutzwiller's trace formula for the oscillatory part of the DOS to graphene cavities with chaotic classical dynamics in \ref{ssec:GTF}.
We conclude in Sec. \ref{sec:conclusion} and gather further technical material in the appendices.

\subsection{Hamiltonian and boundary conditions}

Neglecting the conventional spin degree of freedom, the effective Hamiltonian that describes electron and hole dynamics in graphene close to
half filling is~\cite{Wallace1947}
\begin{equation}
\label{eq:hamiltonian_noniso}
\tilde{H} = \vF\tau_z \otimes \sigma_x \,p_x + \vF\tau_0 \otimes \sigma_y \,p_y\,,
\end{equation}
where $\vF$ is graphene's Fermi velocity. The $\{\sigma_i\}$ denote Pauli matrices in sublattice pseudospin space and Pauli matrices in valley-spin space are repesented by  $\{\tau_i\}$, while $\sigma_0$ and $\tau_0$ are unit matrices acting on the corresponding spin space. In the following, we usually omit the latter. The Hamiltonian (\ref{eq:hamiltonian_noniso}) acts on spinors
$[\psi_A,\psi_B,\psi_A',\psi_{B}']$ where A/B stands for the sublattice index and the primed and unprimed entries correspond to the two valleys. We find it convenient to transform
Eq.\,(\ref{eq:hamiltonian_noniso}) to the valley isotropic form~\cite{Akhmerov2007} using the unitary transformation
\begin{equation}
 \mathcal{U} = \frac{1}{2}(\tau_0+\tau_z)\otimes\sigma_0 + \frac{i}{2}(\tau_0-\tau_z)\sigma_y\,.
\end{equation}
The transformed Hamiltonian is
\begin{equation}
\label{eq:hamiltonian}
H = \mathcal{U}^\dagger \tilde{H} \mathcal{U} = \vF\tau_0 \otimes \boldsymbol{\sigma}\cdotsh\mathbf{p}
\end{equation}
and acts on spinors $[\psi_A,\psi_B,-\psi_B',\psi_{A}]$.

We consider a graphene flake in which electron and hole dynamics is confined to an area $\mathcal{V}$. The boundary
condition on the spinors at a point $\bsy{\alpha}$ on the boundary
$\partial \mathcal{V}$ is expressed as $P_{\bsy{\alpha}} \psi|_{\bsy{\alpha}}= 0$, where $P_{\bsy{\alpha}}$ is a $4\times 4$ projection matrix~\cite{McCann2004, Akhmerov2007}.
Throughout this paper we reserve bold Greek letters for boundary points and bold Roman letters for points in the  bulk of the flake. For the most common boundaries, i.\,e. zigzag (zz), armchair (ac) and infinite mass (im), the boundary matrices are given by\cite{Akhmerov2008}
\begin{equation}
\label{eq:bc}
 P_{\bsy{\alpha}} = \frac{1}{2} \left(1 - \bsy{\nu}\cdotsh\bsy{\tau}\otimes \bsy{\eta}\cdotsh\bsy{\sigma} \right)
\end{equation}
where the vectors $\bsy{\nu}$ and $\bsy{\eta}$ are summarized in Tab. \ref{tab:bc}.
$K=4\pi/3a$ is the distance of the Dirac points from the $\Gamma$-point of the reciprocal space, $x_{\bsy{\alpha}} = \bsy{\alpha}\cdot \bsy{\hat{x}}$ and $\bsy{\hat{t}_\alpha}$ is the direction of the tangent to $\partial \mathcal{V}$ at $\bsy{\alpha}$.
For zigzag edges the sign in $\bsy{\eta}$ is determined by the sublattice of which the zigzag edge consists.
For an $A$-edge the upper sign is valid and for a $B$-edge the lower sign. That means the orientation of the edge
effectively determines $\bsy{\eta}$. 
For armchair edges, the upper sign is valid when the order of the atoms within each dimer is $A$-$B$ along the direction of $\bsy{\hat{t}_\alpha}$, and the lower sign is valid for $B$-$A$ ordering. For infinite mass edges, the sign depends only on the sign of the infinite mass. The upper sign is valid for the mass going to $+\infty$ outside of $\mathcal{V}$ and the lower for the mass going to $-\infty$.

We note that for a model that includes next nearest neighbour hopping (nnn),
the boundary conditions need to be modified to include differential operations on the spinor.
Nevertheless, as we shall show in App.\,\ref{app:ZZNNN}, it is possible to modify our formalism to account for nnn hopping approximately by keeping only nearest neighbor hoppings, but modifying the boundary
conditions introducing an edge potential.
\begin{table}
 \begin{tabular}{c||c|c|c}
 ~ & ~zz~ & ~ac~ & ~im~ \\ \hline\hline
$\bsy{\nu}~$& $\bsy{\hat{z}}$  &~ $-\sin(2Kx_{\bsy{\alpha}})\bsy{\hat{x}}$ ~&  $\bsy{\hat{z}}$ \\
$~$ & $~$ & ~$+\cos(2Kx_{\bsy{\alpha}})\bsy{\hat{y}}$~ & $~$  \\
$\bsy{\eta}~$ &~ $\pm\bsy{\hat{z}}$~ &$ ~\pm \bsy{\hat{t}_\alpha}$ ~&  $~\pm \bsy{\hat{t}_\alpha}~$ \\
\end{tabular}
\caption{The vectors $\bsy{\nu}$ and $\bsy{\eta}$ for zigzag (zz), armchair (ac) and infinite mass (im) type boundaries.}
 \label{tab:bc}
\end{table}

\subsection{Single particle density of states}
The single particle DOS for a closed system is defined as\cite{note_1}
\begin{equation}
 \rho(\kE) = \sum_n \delta\left(\kE - k_n\right)\,.
\end{equation}
Here $n$ labels the eigenenergies $E_n = \hbar \vF k_n$, and we define $E = \hbar \vF \kE$.
In our derivation below we use the relation between the DOS and the retarded Green's function of a system,
\begin{equation}
\label{eq:dos2}
  \rho(\kE)  = -\frac{1}{\pi}\mathfrak{Im} \intl_{\mathcal{V}} \!d\bsy{x} \,\text{Tr}\left[G(\bsy{x},\bsy{x})\right]\,,
\end{equation}
where the Green's function $G$ fulfills
\begin{equation}
\label{eq:Green's_EOM}
(E+i\eta -H)G(\bsy{x},\bsy{x}')=\hbar \vF \delta(\bsy{x}-\bsy{x}')\,,
\end{equation}
with the Hamiltonian $H$ acting on the first argument of $G$. 
For a mesoscopic graphene flake the mean level spacing $\Delta k$, which is given by the inverse area of the system, is typically of the order $10^{-4}\,1/a$ or smaller. This means that $\rho$ is in principle a rapidly oscillating function of $\kE$. 
However, one can decompose $\rho$ into a smooth part $\bar{\rho}$
and an oscillating part $\rho_{\text{osc}}$  in a well defined way\cite{Brack2008, Gutzwiller1990},
\begin{equation}
\label{eq:rho_decomp}
 \rho = \bar{\rho}+\rho_{\text{osc}}\,.
\end{equation}
In this work, we address both contributions to $\rho$ and focus on the
particularities that arise due to
the spinor character and the linear dispersion of quasiparticles in graphene.
The smooth part $\rhobar$ represents the density of states in the limit of strong level
broadening. 
Technically, level broadening is achieved by adding a finite imaginary
part to the Fermi energy or in other words considering a real self energy. This corresponds to an
exponential damping of the Green's function and therefore only trajectories of short length, in the limiting case of `zero-length',
contribute. In Sec. \ref{sec:Weyl} we treat $\rhobar$ in detail.
On the other hand, $\rho_{\text{osc}}$ is connected to (periodic) orbits of finite length, and in
Sec. \ref{sec:OscillatingDOS} we use a semiclassical approach
to describe this part of the density of states.

In the following, we derive an exact expression for the Green's function entering Eq.\,(\ref{eq:dos2}) and later
also its asymptotic form in the semiclassical limit, valid for large system sizes.

\section{The multiple reflection expansion for graphene}
\label{sec:mre}
\begin{figure}
 \centering
 \includegraphics[width=0.58\columnwidth]{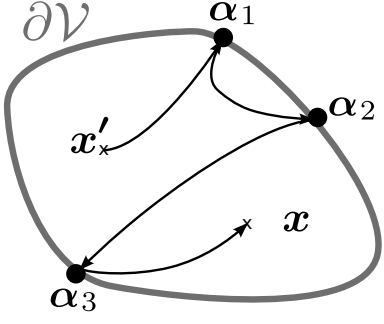}
\caption{
\small{ Schematic representation of a quantum path contributing to the Green's function $G(\bsy{x},\bsy{x}')$. The black lines with arrows stand for free propagations described by $G_0$, while each black disk represents a vertex of the form $i \sigma_{\bsy{n_{\alpha}}} P_{\bsy{\alpha}}$.
  }}
\label{Fig:mre}
\end{figure}
In this chapter, we derive a formula for the exact Green's function of a graphene cavity. The Green's function can then be used to obtain e.\,g. the spectral density of states or the conductance.
In addition to Eq.\,(\ref{eq:Green's_EOM}), $G$ also obeys 
the boundary conditions $P_{\bsy{\alpha}}G(\bsy{\alpha},{\bf x}') = 0$ for any given point $\bsy{\alpha}$ on the boundary.

We now parameterize the full Green's function as a sum of the free retarded Green's function $G_0$ of extended graphene and a boundary correction that is produced by a, yet unknown, Dirac-charge layer $\mu$:
\begin{equation}
\label{eq:FullG1}
 G(\bsy{x}, \bsy{x}') = G_0(\bsy{x}, \bsy{x}')
- \!\!\int\limits_{\partial \mathcal{V}}  \!\!d\sigma_{\bsy{\beta}} \,G_0(\bsy{x}, \bsy{\beta})
 \, \imi {\sigma}_{\bsy{n_\beta}}\,\mu({\bsy{\beta}, \bsy{x}'})\,.
\end{equation}
Here $\sigma_{\bsy{v}} \equiv \bsy{\sigma}\cdotsh \bsy{v}$ for an arbitrary vector $\bsy{v}$, and
$\bsy{n_\beta}$ stands for the normal unit vector at the boundary point $\bsy{\beta}$ pointing into the interior of the system.
The free Green's function is obtained by solving Eq.\,(\ref{eq:Green's_EOM}) with boundary conditions $G_0(\bsy{x}, \bsy{x}') \rightarrow 0$ as $|\bsy{x} - \bsy{x}'| \rightarrow \infty$. It is given by
\begin{eqnarray}
\label{eq:freeG}
  G_0(\bsy{x}, \bsy{x}') &=&  \hbar \vF \bra{\bsy{x}} (E - H)^{-1} \ket{\bsy{x'}} \nonumber \\
&=& -\frac{\imi}{4}(\kE-\imi\bsy{\nabla}_{\bsy{x}}\cdotsh\bsy{\sigma}) H_0^{+}(\kE|\bsy{x}-\bsy{x}'|)\,,
\end{eqnarray}
where $H_0^{+}$ denotes the zeroth order Hankel function of the first kind.  
The free Dirac Green's function can be expressed in terms of the free Schr\"odinger Green's function $g_0$ as
\begin{equation}
  G_0(\bsy{x}, \bsy{x}') = (\kE-\imi\bsy{\nabla}_{\bsy{x}}\cdotsh\bsy{\sigma}) g_0(\bsy{x}, \bsy{x}') \,.
\end{equation}
The Schr\"odinger Green's function $g_0$ is a solution to
\begin{equation}
(\kE^2+i\eta - \hat{p}^2/\hbar^2)g_0(\bsy{x},\bsy{x}')= \delta(\bsy{x}-\bsy{x}')\,.
\end{equation}
The parametrization in Eq.~(\ref{eq:FullG1}) is singular in the limit
$\bsy{x} \rightarrow \bsy{\alpha}$~\cite{Balian1970, Adagideli2002}:
\begin{eqnarray}
\label{eq:jump}
 \lim_{\bsy{x}\rightarrow\bsy{\alpha}} G(\bsy{x}, \bsy{x}') &=& G_0(\bsy{\alpha}, \bsy{x}')
   -\frac{1}{2}\mu({\bsy{\alpha}, \bsy{x}'}) \\
 &-&
 \int\limits_{\partial \mathcal{V}} \!d\sigma_{\bsy{\beta}} G_0(\bsy{\alpha}, \bsy{\beta})
\,\imi \sigma_{\bsy{n_\beta}} \, \mu({\bsy{\beta}, \bsy{x}'})\,. \nonumber 
\end{eqnarray}
The source of this singular behavior is
the logarithmic divergence of $H_0^{+}(\xi)$ as $\xi \rightarrow 0$. For a detailed
derivation of Eq.\,(\ref{eq:jump}) see App.\,\ref{app:jump}. Multiplying (\ref{eq:jump}) with $P_{\bsy{\alpha}}$ and invoking the boundary conditions,
we obtain an inhomogeneous integral equation for the charge layer $\mu$.
As a first step we assume that $P_{\bsy{\alpha}}\mu=\mu$, so that we get
\begin{eqnarray}
\label{eq:Int2}
  \mu({\bsy{\alpha}, \bsy{x}'}) &=& 2 P_{\bsy{\alpha}} G_0(\bsy{\alpha}, \bsy{x}') \\ 
&&- 2 \int\limits_{\partial \mathcal{V}} \!d\sigma_{\bsy{\beta}}\,P_{\bsy{\alpha}} G_0(\bsy{\alpha}, \bsy{\beta})
\, \imi \sigma_{\bsy{n_\beta}}\, \mu({\bsy{\beta}, \bsy{x}'})\,. \nonumber 
\end{eqnarray}
Since $P_{\bsy{\alpha}}^2 = P_{\bsy{\alpha}}$, the unique solution of Eq.\,(\ref{eq:Int2}), obtained by iteration, automatically fullfills $P_{\bsy{\alpha}}\mu=\mu$, and thus is already a solution of the original integral equation for $\mu$.
Substituting this solution into Eq.\,(\ref{eq:FullG1}), we obtain the following expansion for the exact
Green's function of a graphene flake with generic edges:
\begin{eqnarray}
 \label{eq:fullG2_1}
  G(\bsy{x}, \bsy{x}')&=& G_0(\bsy{x}, \bsy{x}') + \sum_{N=1}^{\infty} G_N(\bsy{x}, \bsy{x}')\,.
\end{eqnarray}
where
\begin{eqnarray}
 \label{eq:fullG2_2}
&&\hspace*{-0.cm} G_N(\bsy{x}, \bsy{x}') =  (-2)^N \int\limits_{\partial \mathcal{V}} \!d\sigma_{\bsy{\alpha}_N}\ldots d\sigma_{\bsy{\alpha}_{2}} d\sigma_{\bsy{\alpha}_1} \times \\ 
&& \hspace*{-0.cm} G_0(\bsy{x}, \bsy{\alpha}_N) i \sigma_{\bsy{n_\alpha}_N} P_{\bsy{\alpha}_N} \ldots
G_0(\bsy{\alpha}_2, \bsy{\alpha}_1) i \sigma_{\bsy{n_{\alpha_1}}} P_{\bsy{\alpha}_1}G_0(\bsy{\alpha}_1, \bsy{x}')\,. \nonumber
\end{eqnarray}
Each term in this expansion can be viewed as a sequence of free propagations connected at reflections at the boundary (see Fig.\,\ref{Fig:mre}).
We thus obtain the \textit{multiple reflection expansion} (MRE).
In Eq.\,(\ref{eq:fullG2_2}) every reflection is represented by a boundary dependent projection $P_{\bsy{\alpha}}$ and by $\sigmav{\bsy{n_\alpha}}$, a reflection of the pseudospin across the normal axis given by $\bsy{n_\alpha}$. The integrals along the boundary can be interpreted as a \textquoteleft summation\textquoteright~over all quantum paths leading from $\bsy{x}'$ to $\bsy{x}$. In Fig.\,\ref{Fig:mre}, we show schematically a typical term in the MRE using the example of a quantum path that includes three reflections at the boundary. To summarize at this stage, with Eqs.\,(\ref{eq:fullG2_1}, \ref{eq:fullG2_2}) we obtained a formalism that naturally relates the edge effects to  any quantity that involves single particle Green's functions.

\section{The smoothed density of states of graphene billiards}
\label{sec:Weyl}

\subsection{Weyl expansion}
\begin{figure}
 \centering
 \includegraphics[width=0.86\columnwidth]{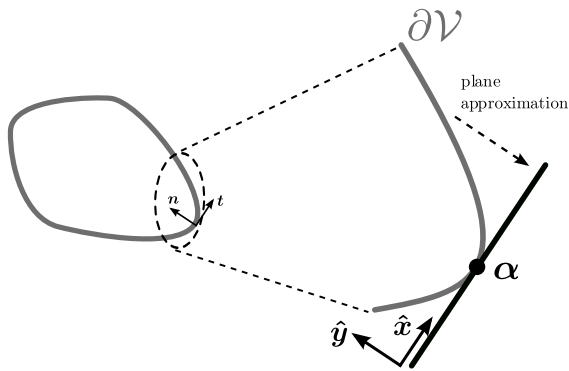}
\caption{
\small{For the calculation of the one-reflection term in the expansion for $\bar{\rho}$, we work in the plane approximation: For a given point $\bsy{\alpha}$ on the boundary $\partial \mathcal{V}$, we approximate the boundary locally by the tangent at $\bsy{\alpha}$ and introduce a local coordinate system with $x$ and $y$ along the tangential and normal direction respectively.
  }}
\label{Fig:plane_approx}
\end{figure}
In the following we are going to derive the leading order contributions to the smoothed density of states $\bar{\rho}$.
In usual Schr\"odinger billiards of linear system size $L$, as they are realized e.\,g. in 2DES in GaAs heterostructures, $\bar{\rho}$ can be expanded in powers of $k_E L$ with leading order $(k_E L) ^1$, a constant term $(k_E L)^0$ and higher order terms $(k_E L)^{-1}$, $(k_E L)^{-2}$ and so forth as
\begin{equation}
\label{eq:WeylEx}
 \bar{\rho} = \bar{\rho}_0 + \bar{\rho}_1 + \bar{\rho}_{2} + \bar{\rho}_{3} \ldots\,.
\end{equation}
In the large $\kE L$ limit, $\bar{\rho}$ is dominated by the first term, which does not depend on the shape of the system but only on its total area. This theorem goes back to Hermann Weyl\cite{Weyl1911} and therefore the series is  known
as the Weyl expansion for the density of states.
Each of the terms in Eq.\,(\ref{eq:WeylEx}) can be obtained from the MRE (\ref{eq:fullG2_2}): $\bar{\rho}_0$ originates from the zero-reflection term (simply $G_0$) and therefore scales with the total area $A$ of the system. The term $\bar{\rho}_1$ is due boundary contributions, obtained within the so-called plane approximation 
(cf. Fig.\,\ref{Fig:plane_approx}), leading to a scaling with the length of the boundary. The term $\bar{\rho}_2$ stems from curvature and corner corrections to the plane approximation and so forth. In this work we focus on leading contributions $\bar{\rho}_0$ and $\bar{\rho}_1$. The smooth contributions
are of qualitatively different origin than the oscillating part of the DOS,
treated in Sec. \ref{sec:OscillatingDOS}. While the latter correspond to orbits for which the phases
occuring in Eq.\,(\ref{eq:dos2}) are stationary,
the smooth DOS is due to trajectories approaching `zero-length' for which the amplitudes diverge.
We find that the linear term in the Weyl expansion for graphene  $\bar{\rho}_0$ is similar to
the usual 2DES case, but the term $\bar{\rho}_1$ behaves strikingly different.

\subsection{Bulk term}

We begin with the zero-reflection term $G_0(\bsy{x},\bsy{x})$ in graphene. From Eq.\,(\ref{eq:freeG})
we can directly read off
\begin{equation}
 \text{Tr}\left[G_0(\bsy{x},\bsy{x'})\right] = -i\kE H_0^+(\kE|\bsy{x}-\bsy{x'}|)\,.
\end{equation}
Although $G_0$ diverges as $\bsy{x'}\rightarrow \bsy{x}$,\cite{Stockmann1999} its imaginary part is finite. We get
\begin{equation}
\mathfrak{Im}\,\tr\left[G_0(\bsy{x},\bsy{x})\right]
= -|\kE|\,.
\end{equation}
Since there is no $\bsy{x}$ dependence left, the spatial integral
in Eq.\,(\ref{eq:dos2}) gives just $A = |\mathcal{V}|$, the area of the billiard,
and we have
\begin{equation}
\label{eq:bulk}
\bar{\rho}_0(\kE) = \frac{A}{\pi} |\kE|\,.
\end{equation}
As for Schr\"odinger billiards, the bulk term (\ref{eq:bulk}) is proportional to the total area of the system.
The energy dependence of $ \bar{\rho}_0$ is however different, since $\kE$ scales linearly with energy in graphene but has a square root dependence in the Schr\"odinger case.

\subsection{Boundary term}
\subsubsection{Plane approximation}
\label{ssec:plane}

As we show below, the boundary term $\bar{\rho}_1$ depends on $\kE$ as well as 
on the boundary length of the system, in a manner distinctly different from that of Schr\"odinger billiards.
In order to evaluate $\bar{\rho}_1$, we assume that the energy has a finite imaginary part $\xi$.
This smoothens the DOS and makes $G_0$ an exponentially decaying function of the distance between $\bsy{x}$ and $\bsy{x'}$.
We start from Eq.\,(\ref{eq:FullG1}), omit the free propagation term that led to $\bar{\rho}_0$, and obtain for the remaining contribution to the smooth DOS
\begin{equation}
\label{eq:deltarho}
 \delta \bar{\rho} = \frac{1}{\pi}\mathfrak{Im}\sum_i \intl_{\partial \mathcal{V}_i} \!d\sigma_{\bsy{\alpha}} \!\!
 \intl_{\mathcal{V}} \!d\bsy{x} \, \text{Tr} \left[ G_0(\bsy{x},\bsy{\alpha}) i \sigma_{\bsy{n_\alpha}} \mu_i(\bsy{\alpha},\bsy{x}) \right]\,.
\end{equation}
Here we replaced the boundary integration by a sum of integrations over boundary pieces $\partial \mathcal{V}_i$, where the boundary condition is constant for each $i$.
Further $\mu_i(\bsy{\alpha},\bsy{x})$ is defined via Eq.\,(\ref{eq:Int2}) with $\bsy{\alpha} \in \partial \mathcal{V}_i$.
Since $G_0$ is short ranged, the dominant contribution to the boundary integral in Eq.\,(\ref{eq:deltarho}) comes from  configurations where $\bsy{x}$ is near the boundary point $\bsy{\alpha}$, and the integral in Eq.\,(\ref{eq:Int2}) is dominated by contributions where $\bsy{\beta}$ is near $\bsy{\bsy{\alpha}}$. Thus we approximate the surface near $\bsy{\alpha}$ by a plane (cf. Fig.\,\ref{Fig:plane_approx}). The corrections to this approximation are of order $1/\kE R$, with the local radius of curvature $R\sim L$, thus of higher order in the Weyl expansion\cite{Balian1970}.
We now take advantage of the homogeneity of the approximate surface at $\bsy{\alpha}$ and use Fourier transformation along the direction of the tangent to the $\partial \mathcal{V}_i$ at $\bsy{\bsy{\alpha}}$, to get for $\delta \bar{\rho} \approx \bar{\rho}_1$
\begin{equation}
\label{eq:deltarho1}
\bar{\rho}_1 = \frac{1}{\pi}\mathfrak{Im}\!\sum_i  |\partial \mathcal{V}_i|
 \int\limits_0^{\infty}\!\! dy_i \!\! \intl_{-\infty}^{\infty}\!\! \frac{dk}{2\pi}\, \text{Tr} \left[ \delta G_i(k,y_i) \right]\,,
\end{equation}
with
\begin{equation}
\label{eq:deltaG}
 \delta G_i(k,y_i)= G_0(k,y_i) i \sigma_{\bsy{n_\alpha}} \mu_i(k,y_i)\,.
\end{equation}
Here $y_i$ is the ordinate of the local coordinate system at $\bsy{\alpha}$ (see Fig.\,\ref{Fig:plane_approx}) and 
\begin{eqnarray}
\label{eq:mu(k)}
 \mu_i(k,y_i)\! &=&\! 2\Gamma_i(k) P_{\bsy{\alpha}}\, G_0(k,-y_i) \,, \\
 \label{eq:mu(k)_2}
 \Gamma_i(k) &=& \left[1\!+\!2P_{\bsy{\alpha}}\,G_0(k,0) \,\imi\sigma_{y}\right]^{-1}\,,
\end{eqnarray}
with the Fourier transform defined as 
\begin{equation}
f(x,y) = \intl_{-\infty}^\infty \! \frac{dk}{2\pi} \,e^{ikx} f(k,y)\,.
\end{equation}
We pushed the upper limits of the $y_i$-integration to infinity, which is valid when $\exp[-\mathfrak{Im}(\kE) L] \ll 1$. To obtain Eq.\,(\ref{eq:deltarho1}), we further assumed that $\bsy{\alpha}$ is away from the corners where the boundary condition changes. The corrections due to such points are of order $1/\kE L$ smaller than the boundary term.

The free Green's function in mixed representation is given by
\begin{equation}
\label{eq:freeGk}
 G_0(k,y_i) = \frac{-e^{-a(k)|y_i|} }{2a(k)}\left[k\sigma_x+i\,\text{sgn}(y_i)a(k)\sigma_y+\kE\right]
\end{equation}
with
\begin{equation}
 a(k) = \sqrt{k^2-\kE^2},\qquad\mathfrak{Re}[a(k)] > 0\,.
\end{equation}
Next we focus on contributions to the boundary term from various types of edges.

\subsubsection{Zigzag edge}
\label{sssec:Weylzz}

For a zigzag edge (without nnn hopping, see Tab. \ref{tab:bc})
\begin{equation}
\label{eq:bczz_def}
 P_{\bsy{\alpha}} = (1\mp\tau_z\otimes\sigma_z)/2\,.
\end{equation}
Then $\Gamma_i$ is diagonal in valley space
and we can invert the valley subblocks separately giving
\begin{equation}
\label{eq:Gamma_zz}
\Gamma_i(k)
=-\frac{a(k)\pm k\tau_z}{\kE^2} \left[{a(k)-(k\sigma_z-i\kE\sigma_y)(1-P_{\bsy{\alpha}})}\right]\,.
\end{equation}
We insert $\Gamma_i(k)$, Eq.\,(\ref{eq:Gamma_zz}), into Eq.\,(\ref{eq:mu(k)}) and take into account that $P_{\bsy{\alpha}}$ is a projection matrix, i.\,e. 
\mbox{$P_{\bsy{\alpha}}^2=P_{\bsy{\alpha}}$}, to obtain for the Dirac-charge density
\begin{eqnarray}
\label{eq:mu(k)renorm}
 \mu_i(k,y_i) = -2\frac{a(k)}{\kE^2}\left[a(k)\pm k\tau_z\right] P_{\bsy{\alpha}}\, G_0(k,-y_i)\,.
\end{eqnarray}
Substituting this expression into Eq.\,(\ref{eq:deltaG}), we obtain 
\begin{eqnarray}
\label{eq:G1zz}
 &\delta G_i(k,y_i)& \\  &=& \hspace*{-0.4cm} -2\frac{a(k)}{\kE^2}\left[a(k)\pm k\tau_z\right] G_0(k,y_i)\,i\sigma_y \,P_{\bsy{\alpha}}\, G_0(k,-y_i)\,. \nonumber
\end{eqnarray}
Then the trace is given by (note that \mbox{$y_i>0$})
\begin{equation}
 \tr\left[\delta G_i(k,y_i)\right] = -\frac{2k^2}{a(k)\kE} e^{-2a(k)y_i}\,.
\end{equation}
Evaluating the $y_i$-integral we get (note that the real part of $a(k)$ is positive) 
\begin{equation}
\label{eq:calcSDOS_NN}
 \mathfrak{Im} \int\limits_0^{\infty}\!\! dy_i \! \int \frac{dk}{2\pi}\, \tr\left[\delta G_i(k,y_i)\right]
= \kmax \delta_\xi(\kE)\,,
\end{equation}
where
\begin{equation}
 \delta_\xi(\kE) = \frac{1}{\pi}\frac{\xi}{\xi^2+\kE^2}\,,
\end{equation}
and we have introduced a cut-off momentum $k_\text{max}\sim 1/a$. Such a cut-off is justified,
since in real graphene the available $k$-space is not infinite owing to the lattice structure. We cannot calculate the precise numerical value for $\kmax$ within our effective model. Using tight-binding calculations we estimate $\kmax = \pi/3a$\,\cite{Wimmer2008a}.
The result (\ref{eq:calcSDOS_NN}) means that without nnn hopping, zigzag edges lead to a DOS contribution that is strongly peaked at zero
energy. The origin of this contribution is indeed the existence of zigzag edge states at zero energy\cite{Fujita1996, Nakada1996, Wimmer2010, Kobayashi2005, Niimi2006}.
To understand this connection we consider the prefactors in Eq.\,(\ref{eq:mu(k)renorm}) and Eq.\,(\ref{eq:G1zz}) in the limit of $\kE\rightarrow 0$; then
we have
\begin{equation}
\frac{a(k)}{\kE^2}\left[a(k)\pm k\tau_z\right] \approx \frac{k^2}{\kE^2} [1\pm \sgn{(k)}\,\tau_z]\,.
\end{equation}
For the upper sign, this expression is divergent in one valley for negative $k$ ($\tau=+1$) and in the other valley for positive $k$ ($\tau=-1$) as $\kE$ approaches zero.
For the lower sign it is just vice versa. Thus we identify the zero-energy states that are localized at the zigzag graphene edge. In a single valley this causes a strong asymmetry in the spectrum and breaks the (effective) time reversal symmetry.
Below we show that the zigzag edge states are the only contribution to the DOS that scales with the boundary length of the graphene flake. Armchair and infinite mass type edges do not contribute to the surface term. However for the zigzag edge states, the effect of nnn hopping is significant\cite{Wimmer2008a, Sasaki2009, Wimmer2010}. For a more realistic description of the their effects on the DOS, it is therefore necessary to consider nnn hopping for the boundary term at zigzag edges. In App.\,\ref{app:ZZNNN} we show that the boundary condition for zigzag edges is effectively modified due to nnn hopping resulting in a boundary matrix
\begin{equation}
\label{eq:Pnnn}
 P_{\bsy{\alpha}} = \frac{1}{2}\left(1\mp \tau_z \otimes \sigma_z - i t' \sigma_y \pm t' \tau_z \otimes \sigma_x \right)\,.
\end{equation}
Here $t'\ll 1$ is the ratio of the nnn hopping integral and the nearest neighbor hopping integral in the tight-binding formalism.
The effect of this boundary condition is to modify Eq.\,(\ref{eq:mu(k)renorm}) to
\begin{eqnarray}
\label{eq:munnn}
 \mu(k,y') = 2a(k)\frac{a(k)-t'\kE \pm k \tau_z}{[a(k)-t'\kE ]^2- k^2} P_{\bsy{\alpha}}\, G_0(k,-y')\,. \nonumber \\
\end{eqnarray}
Note that the Eqs.\,(\ref{eq:Pnnn}, \ref{eq:munnn}) turn into the expressions (\ref{eq:bczz_def}, \ref{eq:mu(k)renorm}) for $t'=0$. Following the same line of calculation we find
\begin{equation}
\tr\left[\delta G_i(k,y_i)\right] = \frac{2k^2 }{a(k)}\frac{t'^2-1}{(1-t'^2)\kE + 2t' a(k)} e^{-2a(k)y_i}  \! \!\!
\end{equation}
and the corresponding contribution to the DOS is to linear order in $t'$
\begin{equation}
\label{eq:boundaryNNN}
  \mathfrak{Im} \!\!\int\limits_0^{\infty}\!\! dy_i \!\intl_{-\infty}^{\infty} \frac{dk}{2\pi}\, \tr\left[\delta G_i(k,y_i)\right] 
\approx \frac{1-\Theta_\xi(\kE)}{2t'}\,.
\end{equation}
Here
\begin{equation}
 \Theta_\xi(\kE) = \frac{1}{\pi} \arctan(\kE/\xi) + \frac{1}{2}
\end{equation}
is a smooth approximation to the Heaviside step function.
 
According to Eq.\,(\ref{eq:boundaryNNN}), the $\kE$-dependence of the zigzag contribution to the DOS is qualitatively altered
by the inclusion of nnn hopping. It is strongly asymmetric due to the broken electron-hole symmetry\cite{note_2}.
Also the peak at zero $\kE=0$ has disappeared, because the edge states are not degenerate anymore but exhibit a linear dispersion $\kE^{\text{edge}} = k/2t'$ as derived in App. \ref{app:ZZNNN}. Note that in tight-binding, there is still a van Hove singularity in the DOS, but it is at a distance to the $K/K'$ points and therefore
not captured by the effective theory.

\subsubsection{Armchair edge}

We now proceed with armchair type edges.
According to Tab.\,\ref{tab:bc}, the boundary projection matrix is given by
\begin{equation}
 P_{\bsy{\alpha}} = \frac{1}{2}\left(1-\sigma_x\otimes\tau_y\right)\,.
\end{equation}
Then we obtain
\begin{equation}
\label{eq:Gamma_ac}
 \Gamma_i(k)
=  1+ \frac{i}{a(k)}(\kE\sigma_y+ik\sigma_z)(1-P_{\bsy{\alpha}})
\end{equation}
and the surface Dirac-charge density reads
\begin{eqnarray}
 \mu_i(k,y_i) = 2 P_{\bsy{\alpha}}\, G_0(k,-y_i)\,,
\end{eqnarray}
leading to [cf. Eq.\,(\ref{eq:deltaG})]
\begin{eqnarray}
\delta G_i(k,y_i) &=& 2 G_0(k,y_i)\,i\sigma_y \,P_{\bsy{\alpha}}\, G_0(k,-y_i) \nonumber \\
&=& -G_0(k,y_i)\sigma_z G_0(k,-y_i) \otimes \tau_y\,.
\end{eqnarray}
Surprisingly, since $\tau_y$ is off-diagonal, the trace of $\delta G_i$ is zero and the boundary contribution to $\bar{\rho}$ in the armchair case vanishes.

\subsubsection{Infinite mass edge}

The calculation for the infinite mass edge is similar and for the surface Dirac-charge density we find as for the armchair case
\begin{eqnarray}
 \mu_i(k,y_i) = 2 P_{\bsy{\alpha}}\, G_0(k,-y_i)\,,
\end{eqnarray}
which leads to
\begin{equation}
 \delta G_i(k,y_i) = \pm G_0(k,y_i) \sigma_z G_0(k,-y_i) \otimes \tau_z\,.
\end{equation}
Similar as for the armchair edge, this expression is traceless because $\text{Tr}\,(\tau_z) =0$. However,
we point out that even within individual valleys the boundary contribution to the DOS vanishes.
This follows from the fact that
\begin{equation}
 \intl_0^{\infty}\!dy_i \,\tr\left[G_0(k,y_i) \sigma_z G_0(k,-y_i)\right] \sim \frac{k}{a^2(k)}
\end{equation}
is an odd function of $k$ and thus the corresponding integral vanishes. This last fact has been already noticed by Berry and Mondragon\cite{Berry1987} for massless neutrinos in relativistic billiards with infinite mass walls.

\subsection{Comparison with numerical results for various graphene billiards}
\begin{figure}
 \centering
  \includegraphics[width=0.45\textwidth]{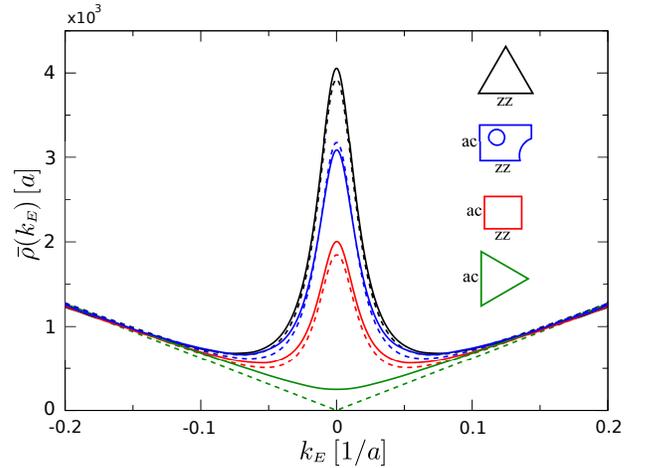}
\caption{\small{(color online).
Smooth part of the density of states for several graphene billiards with approximately the same area $A \approx (140\,a)^2$, calculated numerically using a tight-binding code with only nearest neighbor coupling (solid lines). The numerical curves are obtained by first calculating exact eigenenergies and successive smoothing by replacing each energy level by a Lorentzian with a half width at half maximum of $0.015\,t$. The dashed lines are the predictions of our theory, Eq.\,(\ref{eq:SDOS_NN}). From top to bottom:
\mbox{black: $|\partial\mathcal{V}_{zz}|/|\partial\mathcal{V}| = 1$} (zigzag triangle), 
\mbox{blue: $|\partial\mathcal{V}_{zz}|/|\partial\mathcal{V}| \approx 1/1.6$} (Sinai shape), 	
\mbox{red: $|\partial\mathcal{V}_{zz}|/|\partial\mathcal{V}| \approx 1/1.9$} (rectangle),
\mbox{green: $|\partial\mathcal{V}_{zz}|/|\partial\mathcal{V}| = 0$} (armchair triangle).
}}
\label{Fig:WeylNN}
\end{figure}
\begin{figure}
 \centering
 \includegraphics[width=0.45\textwidth]{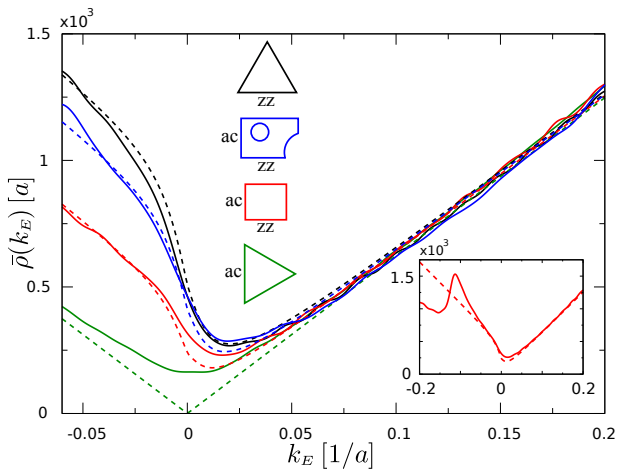}
\caption{\small{(color online).
Smooth part of the density of states for the same systems as in Fig.\,\ref{Fig:WeylNN} but with a relative next-nearest neighbor hopping strength $t'=0.1$.
Solid lines show the numerical tight-binding results and dashed lines the predictions from Eq.\,(\ref{eq:SDOS_NNN}). For the smoothing we used Lorentzians with a half width at half maximum of $0.01\,t$. We used the same color coding as for Fig.\,\ref{Fig:WeylNN}.
Inset: The tight-binding model exhibits a van Hove singularity at $\kE = -0.1\,t/\hbar\vF \approx -0.115\,1/a$. 
As a result the smoothed DOS shows a peak at the corresponding position (solid). }}
\label{Fig:WeylNNN}
\end{figure}
In summary, our result for the smooth DOS of a generic graphene billiard, neglecting the effect of next-nearest neighbors is
\begin{equation}
\label{eq:SDOS_NN}
 \bar{\rho}(\kE) \approx \frac{A}{\pi} |\kE| + |\partial\mathcal{V}_{zz}| \frac{k_{\text{max}}}{\pi}\delta_\xi(\kE) \,,
\end{equation}
with $|\partial\mathcal{V}_{zz}| $ being the total length of zigzag edges in the billiard. 
 
In Fig.\,\ref{Fig:WeylNN} we compare our analytical result (\ref{eq:SDOS_NN}) with results from
numerical simulations for the graphene billiards shown as insets. For the numerical calculations we obtain the average DOS by computing eigenvalues of a corresponding tight-binding Hamiltonian\cite{Wimmer2009, Wimmer2010} and subsequent smoothing. All the billiards are chosen to have approximately the same area. This is reflected in the common slope of $\bar{\rho}$ for larger $\kE$, confirming the leading order term in the Weyl series. The different shapes and orientations give rise to different fractions of the zigzag boundary $|\partial\mathcal{V}_{zz}|/|\partial\mathcal{V}|$. While the boundaries of the equilateral triangles consist
completely of either zigzag (black) or armchair (green) edges, both edge types are present in the rectangle (red) and in the non-integrable (modified) Sinai billiard (blue).  We find very good agreement with our analytic prediction. We note that the dashed lines for the triangles and the rectangle do not involve any fitting, rather we have used the estimation $\kmax = \pi/3a$ from tight-binding theory. For the Sinai billiard our theory allows to determine the total effective zigzag length $|\partial\mathcal{V}_{zz}| = 516\,a$.

On the other hand, with nnn hopping we get from Eq.\,(\ref{eq:boundaryNNN})
\begin{equation}
\label{eq:SDOS_NNN}
 \bar{\rho}(\kE) \approx \frac{A}{\pi} |\kE| + |\partial \mathcal{V}_{zz}| \frac{1-\Theta_\xi(\kE)}{2\pi t'}\,.
\end{equation}
In Fig.\,\ref{Fig:WeylNNN} we compare again this analytical result (dashed) with corresponding tight-binding calculations (solid). 
Also here we find good agreement with our analytic predicition for the surface term. Further towards the hole regime, i.\,e. to more negative energies, the tight-binding model has a van Hove singularity due to the edge state band edge at $\kE  = -0.1\,t/\hbar\vF \approx -0.115\,1/a$, as depicted in the inset of Fig.\,\ref{Fig:WeylNNN} (solid line). This peak is missing in our calculation, since in the effective Dirac theory the edge state dispersion is constantly linear for finite $t'$ (cf. App. \ref{app:ZZNNN}). 
Note that also here, no additional fitting is involved (for the Sinai billiard we use $|\partial \mathcal{V}_{zz}| = 516\,a$ obtained from the fit in  Fig.\,\ref{Fig:WeylNN}). 

From our discussion in this section it becomes clear that in principle the structure of a
graphene flake's boundary, i.\,e. the ratio between zigzag and armchair type edges, can be estimated from the behavior of the smoothed density of states at low energies.
Hereby the formula (\ref{eq:SDOS_NN}) predicts the spectral weight of the edge states $\int_{-\infty}^{\infty} d\kE \, \bar{\rho}_1(\kE) = |\partial \mathcal{V}_{zz}|/3a$, which is model independent, since the number of edge states is conserved. 
 Note that Libisch \textit{et al.} have numerically investigated\cite{Libisch2009} the averaged DOS of graphene billiards and found a $\bar{\rho}(\kE)$ profile similar to that in Fig.\,\ref{Fig:WeylNN}. Related studies on edge states in graphene quantum dots have been performed in Ref. \onlinecite{Wimmer2010}.

\section{Density of states oscillations}
\label{sec:OscillatingDOS}

\subsection{The multiple reflection expansion in the semiclassical limit}
\label{ssec:Semiclassics}

So far we have focused on the smooth part of the density of states. In this section we study the
oscillating part $\rhoosc$. 
Our main result is an extension of Gutzwiller's trace formula\cite{Gutzwiller1990} to graphene systems with chaotic and regular classcial dynamics.
We derive the trace formulae by evaluating Eq.\,(\ref{eq:dos2}) asymptotically in the semiclassical limit $\kE L \gg 1$. In other words we evaluate the boundary integrals in the MRE (\ref{eq:fullG2_2}) using the method of stationary phase.
In the limit $\kE L \gg 1$, the Hankel functions become rapidly oscillating exponential functions of the boundary points. All other terms in
$G_N$ vary slowly  along $\partial \mathcal{V}$. Thus we evaluate them at the critical boundary points where the total phase of the exponentials is stationary.
There is another leading-order contribution to the boundary integrals that is of different origin,
namely when the set of boundary points $\underline{\bsy{\alpha}} = (\bsy{\alpha}_N, \ldots, \bsy{\alpha}_1)$ leads to a singularity in the prefactors \cite{Balian1972, Adagideli2002}.
Due to the divergence of $G_0(\bsy{\alpha},\bsy{\beta})$ as $|\bsy{\alpha}-\bsy{\beta}|\rightarrow 0$, quantum paths involving reflections at closely lying
boundary points can give rise to such singularities. We show below that short range critical points occur only at zigzag edges.
We treat these short range singularities at zigzag edges by resumming the MRE leading to a renormalized reflection operator.

\subsubsection{Resummation of short range processes}
\label{ssec:resumm}
\begin{figure}
 \centering
 \includegraphics[width=0.25\textwidth]{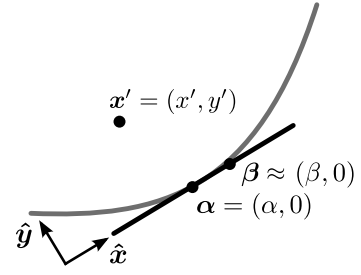}
\caption{\small{Notation in the local coordinate system spanned by the tangent and the normal to the boundary at $\bsy{\alpha}$. Corrections to the approximation $\bsy{\beta} \approx (\beta, 0)$ are of subleading order in $\kE L$, cf. \ref{ssec:plane}. }}
\label{Fig:plane_approx_2}
\end{figure}

The general method is outlined in Ref.\,\onlinecite{Adagideli2002}. Here we apply it to graphene. 
First we isolate the short range singularities: We define the action of an operator $\OpI$ on a function $f$
\begin{equation}
\OpI f(\bsy{\alpha}):= \int\limits_{\partial V }\! d\sigmav{\bsy{\beta}} \, \I(\bsy{\alpha}, \bsy{\beta}) f(\bsy{\beta}) \,.
\end{equation}
In our case
\begin{equation}
 \I(\bsy{\alpha}, \bsy{\beta}) = 2 P_{\bsy{\alpha}} G_0(\bsy{\alpha}, \bsy{\beta})\, i\sigmav{\bsy{n_{\beta}}}\,.
\end{equation}
We now recast Eq.\,(\ref{eq:Int2}) as
\begin{equation}
\label{eq:Int3}
   \mu({\bsy{\alpha}, \bsy{x}'}) = 2 P_{\bsy{\alpha}} G_0(\bsy{\alpha}, \bsy{x}')
- \OpI \mu({\bsy{\alpha}, \bsy{x}'})\,.
\end{equation}
Furthermore we decompose $\I$ into a short range part $\Is$ and a long range part $\Il$:
\begin{equation}
\begin{split}
 &  \Is(\bsy{\alpha},\bsy{\beta}) = \I(\bsy{\alpha},\bsy{\beta})\left[1-w(\bsy{\alpha}-\bsy{\beta})\right]\,, \\
 &  \Il(\bsy{\alpha},\bsy{\beta}) = \I(\bsy{\alpha},\bsy{\beta})  w(\bsy{\alpha}-\bsy{\beta})\,.
\end{split}
\end{equation}
Here $w(\bsy{\alpha}-\bsy{\beta})$ is a smooth function, that is zero whenever $\bsy{\alpha}$ is close to $\bsy{\beta}$ and goes to one otherwise, so that
integrating over $\bsy{\beta}$ isolates the critical point $\bsy{\beta}=\bsy{\alpha}$. This separation is however a formal one in that the specific form
of $w$ does not change the final result (see Ref.\,\onlinecite{Bleistein1975} for details). 
Then Eq.\,(\ref{eq:Int3}) leads to
\begin{equation}
 (1+\OpIs)\,\mu({\bsy{\alpha}, \bsy{x}'}) = 2 P_{\bsy{\alpha}} G_0(\bsy{\alpha}, \bsy{x}')
- \OpIl\, \mu({\bsy{\alpha}, \bsy{x}'})
\end{equation}
or with $\hat{\Gamma} = (\hat{1}+\OpIs)^{-1}$

\begin{equation}
 \mu({\bsy{\alpha}, \bsy{x}'}) = 2\hat{\Gamma} P_{\bsy{\alpha}} G_0(\bsy{\alpha}, \bsy{x}')
- \hat{\Gamma}\, \OpIl\, \mu({\bsy{\alpha}, \bsy{x}'})\,.
\end{equation}
Now the renormalized Kernel $\OpIl$ is free of short range singularities.
Alternatively, in integral representation
\begin{eqnarray}
\label{eq:IntRenorm}
  \mu({\bsy{\alpha}, \bsy{x}'}) &=& 2\int\limits_{\partial V} d\sigma_{\bsy{\beta}}\,\Gamma({\bsy{\alpha}, \bsy{\beta}}) P_{\bsy{\beta}} G_0(\bsy{\beta}, \bsy{x'})
 \\
&& - \int\limits_{\partial V} d\sigma_{\bsy{\beta}}\,\int\limits_{\partial V} d\sigma_{\bsy{\beta}'}\,\Gamma({\bsy{\alpha}, \bsy{\beta}}) {\mathcal{I}}_{\text{l}}({\bsy{\beta}, \bsy{\beta}'})\, \mu({\bsy{\beta}', \bsy{x}'})\,. \nonumber
\end{eqnarray}
We note that the relevant structure of both terms in this expression is the same, since $\mathcal{I}_{\text{l}}$ contains the isolating function $w$ and thus $\bsy{\beta}'$ can be considered to lie far away from $\bsy{\beta}$ just as $\bsy{x}'$ in the first term. In this way we have formally collected all the short range contributions in $\Gamma$ and we are left with calculating
\begin{equation}
\label{eq:PGRenorm}
2 \int\limits_{\partial V} \!d\sigma_{\bsy{\beta}}\,\Gamma({\bsy{\alpha}, \bsy{\beta}}) P_{\bsy{\beta}} G_0(\bsy{\beta}, \bsy{x'})\,.
\end{equation}
We evaluate Eq.\,(\ref{eq:PGRenorm}) again in the plane approximation and replace the boundary in the vicinity of $\bsy{\alpha}$ by a straight line in the direction of the tangent at $\bsy{\alpha}$. In our local coordinate system with $x$ and $y$ denoting coordinates in the tangential and normal directions, we approximate a point $\bsy{\beta}$ close to $\bsy{\alpha}$ by $\bsy{\beta} = (x_{\bsy{\beta}},y_{\bsy{\beta}}) \approx (\beta, 0)$, and write $\bsy{x'} = (x',y')$ for a point $\bsy{x'}$ far away from $\bsy{\alpha}$ (cf. Fig.\,\ref{Fig:plane_approx_2}).  Then the system is locally homogeneous along the straight boundary and we have
\begin{eqnarray}
 \Gamma({\bsy{\alpha}, \bsy{\beta}}) &=& \Gamma(\alpha-\beta)\,, \\ G_0(\bsy{\beta}, \bsy{x'}) &=& G_0(\beta-x', -y')\,.
\end{eqnarray}
In order to partial Fourier transform the expression (\ref{eq:PGRenorm}), we use the convolution theorem to obtain ($P_{\bsy{\alpha}}=P_{\bsy{\beta}}$ is
constant along the straight boundary)
\begin{eqnarray}
\label{eq:PGRenormk}
 \intl_{-\infty}^\infty \! d\beta\,&& \hspace*{-0.45cm}\Gamma({{\alpha}-{\beta}}) P_{\bsy{\alpha}}  G_0({\beta}-{x}',-y')  \nonumber \\
&=& \intl_{-\infty}^\infty \!dk \,e^{ik({\alpha}-{x'})} \Gamma(k) P_{\bsy{\alpha}} G_0(k,-y') \,.
\end{eqnarray}
In fact we have calculated $\Gamma(k)$ already earlier, cf. Eq.\,(\ref{eq:Gamma_zz}) and Eq.\,(\ref{eq:Gamma_ac}), leading to 
\begin{equation}
 \Gamma(k) P_{\bsy{\alpha}} = R_{\bsy{\alpha}}(k) P_{\bsy{\alpha}}
\end{equation}
with the renormalizing factor
\begin{equation}
\label{eq:Ralpha_1}
R_{\bsy{\alpha}}(k) =
 \left\{
\begin{array}{cl}
 -\frac{a(k)}{\kE^2}\left[a(k)\pm k\tau_z\right] & \text{for~zz~edges}\,, \\
1 & \text{for~ac~and~im~edges}\,.
\end{array}
\right. 
\end{equation}
We now define the renormalized free Green's function through its Fourier transform as
\begin{eqnarray}
\label{eq:RenormG}
\tilde{G}_0(\bsy{\alpha},\bsy{x'}) &=& \intl_{-\infty}^\infty \frac{dk}{2\pi} \,e^{ik({\alpha}-{x'})}R_{\bsy{\alpha}}(k) G_0(k,-y')\,.
\end{eqnarray}
Finally we cast Eq.\,(\ref{eq:PGRenorm}) for the charge layer $\mu$ in position space into the form 
\begin{eqnarray}
\label{eq:Renormmu}
  \mu({\bsy{\alpha}, \bsy{x}'}) &=& 2 P_{\bsy{\alpha}} \tilde{G}_0(\bsy{\alpha}, \bsy{x}') \\
&-& 2 \int\limits_{\partial \mathcal{V}} \!d\sigma_{\bsy{\beta}}\,P_{\bsy{\alpha}} \tilde{G}_0(\bsy{\alpha}, \bsy{\beta}) w(\bsy{\alpha}- \bsy{\beta})
\, \imi \sigma_{\bsy{n_\beta}}\, \mu({\bsy{\beta}, \bsy{x}'}) \nonumber.
\end{eqnarray}
The virtue of this equation is that it is free of short range singularities.
\subsubsection{Renormalized Green's function in the semiclassical limit}
With the definition
\begin{equation}
\label{eq:deftheta}
 \theta(k) = \arctan\left(\frac{k}{\sqrt{\kE^2-k^2}}\right)
\end{equation}
we obtain from Eq.\,(\ref{eq:Ralpha_1})
\begin{equation}
\label{eq:Ralpha_2}
R_{\bsy{\alpha}}(k) =
 \left\{
\begin{array}{cl}
\cos[\theta(k)]\,e^{\pm i\theta(k)\tau_z}  & \text{for~zz~edges}\,, \\
1 & \text{for~ac~and~im~edges}\,.
\end{array}
\right. 
\end{equation}
We compute $ \tilde{G}_0(\bsy{\alpha},\bsy{x'})$ in Eq.\,(\ref{eq:Renormmu}) by performing the Fourier integral Eq.\,(\ref{eq:RenormG}) [with $R_{\bsy{\alpha}}$ from Eq.\,(\ref{eq:Ralpha_2})] within stationary phase approximation in the limit $\kE L \rightarrow \infty$.
We obtain the stationary phase point $k_0$ from
\begin{equation}
\frac{d}{d k} \left[k(\alpha-x')-\sqrt{\kE^2-k^2}\,|y'|\right]_{k_0} =0
\end{equation}
yielding, in view of Eq.\,(\ref{eq:deftheta}),
\begin{equation}
\tan[\theta(k_0)] = \frac{\alpha - x'}{|y'|}\,.
\end{equation}
The stationary phase point $k_0$ is such that the angle $\theta(k_0)$ is equal to the angle that the vector $\bsy{x'} - \bsy{\alpha}$ includes with the normal at $\bsy{\alpha}$, i.\,e. the classical angle of incidence. The stationary phase integration yields
\begin{equation}
\label{eq:RenormG2}
\tilde{G}_0(\bsy{\alpha}, \bsy{x'}) \approx R_{\bsy{\alpha}}(k_0) G_0^{\text{sc}}(\bsy{\alpha}, \bsy{x'})\,.
\end{equation}
Here $G_0^{\text{sc}}$ is the free Green's function in the semiclassical limit
\begin{equation}
\label{eq:freeGsc}
 G_0^{\text{sc}}(\bsy{\alpha}, \bsy{\beta}) = - \frac{i}{4} \sqrt{\frac{2\kE}{\pi |\bsy{\alpha}-\bsy{\beta}|}}e^{ i\kE|\bsy{\alpha}-\bsy{\beta}|- i\pi/4} \left(1+\sigmav{\bsy{\alpha},\bsy{\beta}}\right)
\end{equation}
where we use the short notation \mbox{$\sigmav{\bsy{\alpha},\bsy{\beta}} = \bsy{\sigma} \cdot (\bsy{\alpha}-\bsy{\beta})/|\bsy{\alpha}-\bsy{\beta}|$} in Eq.\,(\ref{eq:freeGsc}). 
We note that expression (\ref{eq:freeGsc}) is closely related to the semiclassical Green's function for the free Schr\"odinger equation $g_0^\text{sc}$, namely
\begin{equation}
 G_0^\text{sc}(\bsy{\alpha}, \bsy{\beta})  = \kE g_0^\text{sc}(\bsy{\alpha}, \bsy{\beta})  \left(1+\sigmav{\bsy{\alpha},\bsy{\beta}}\right)\,.
\end{equation}
The matrix term reflects the chirality of the charge carriers in graphene: the sublattice pseudospin is tied to the propagation direction and the projection
$\left(1+\sigmav{\bsy{\alpha},\bsy{\beta}}\right)$ takes care of this.
Eq.\,(\ref{eq:RenormG2}) together with Eq.\,(\ref{eq:Renormmu}) completes our discussion of the short range divergencies and allows us to proceed with the long range contributions to the Green's function in the semiclassical limit.

\subsubsection{Semiclassical Green's function for graphene cavities}
\label{ssec:Semiclassics1}
In this section we evaluate the boundary integrals in the renormalized MRE in stationary phase approximation. We consider the $N$-reflection term [cf. Eq.\,(\ref{eq:fullG2_2})] of the renormalized MRE,
\begin{eqnarray}
\label{eq:GN_1}
G_N(\bsy{x},\bsy{x'}) &\approx&(-2)^{N}\prod_{i=1}^N \int\limits_{\partial \mathcal{V}} \!d\sigma_{\bsy{\alpha}_i} \tilde{K}_N(\underline{\bsy{\alpha}})\, \kE g_0^\text{sc}(\bsy{x}, \bsy{\alpha}_N) \nonumber \\
 &&\ldots i\kE g_0^\text{sc}(\bsy{\alpha}_2, \bsy{\alpha}_1) \,i\kE g_0^\text{sc}(\bsy{\alpha}_1, \bsy{x'})\,,
\end{eqnarray}
with $\underline{\bsy{\alpha}}=(\bsy{\alpha}_1, ..\bsy{\alpha}_i, ..\bsy{\alpha}_N)$. In Eq.\,(\ref{eq:GN_1}) we introduced the \textit{pseudospin propagator} $\tilde{K}_N(\underline{\bsy{\alpha}})$ that contains the graphene specific physics:
\begin{eqnarray}
\label{eq:pseudospinprop}
 \tilde{K}_N(\underline{\bsy{\alpha}}) &=& \left(1+\sigmav{\bsy{x},\bsy{\alpha}_N}\right) \prod_{i=1}^{N-1}  \sigmav{\bsy{n}_{\bsy{\alpha}_i}}
 {R}_{\bsy{\alpha}_i} P_{\bsy{\alpha}_i}
\left(1+\sigmav{\bsy{\alpha}_{i+1},\bsy{\alpha}_i}\right) \nonumber \\
&&\times  \sigmav{\bsy{n}_{\bsy{\alpha}_1}} P_{\bsy{\alpha}_1} \left(1+\sigmav{\bsy{\alpha}_{1},\bsy{x'}}\right)  W(\underline{\bsy{\alpha}})
\end{eqnarray}
with the separation function
\begin{equation}
  W(\underline{\bsy{\alpha}}) = \prod_{i=1}^{N-1} w(\bsy{\alpha}_{i+1}-\bsy{\alpha}_{i})\,.
\end{equation}
Note that the renormalization matrices ${R}_{\bsy{\alpha}_i}$ account for possible short range singularities. 

Comparing Eq.\,(\ref{eq:GN_1}) with the MRE for the Helmoltz equation with Dirichlet boundary conditions \cite{Balian1970}
shows that the scalar parts are very similar. The difference is that instead of factors
$i\kE g_0^\text{sc}(\bsy{\alpha}_{i+1}, \bsy{\alpha}_i)$, the MRE in Ref. \onlinecite{Balian1970} has normal derivatives acting on the first argument $\bsy{\alpha}_{i+1}$.
In the semiclassical limit this leads to additional factors $i\kE \cos(\theta_{i+1})$, where $\theta_{i+1}$ denotes the angle between the vector $\bsy{\alpha}_{i+1} - \bsy{\alpha}_{i}$ and the normal vector to the boundary at $\bsy{\alpha}_{i+1}$. We need not carry out the boundary integrals explicitly, but can immediately deduce
\begin{equation}
\label{eq:GN_2}
G^\text{sc} _N(\bsy{x},\bsy{x'})= \kE {K}_N\, g_N^\text{sc}(\bsy{x}, \bsy{x'})\,,
\end{equation}
where
\begin{equation}
\label{eq:Kgamma}
 K_N =  \frac{\tilde{K}_{N}(\underline{\bsy{\alpha}})}{\prod_{i=1}^N \cos(\theta_i)}
\end{equation}
contains the pseudospin propagator as defined in Eq.\,(\ref{eq:pseudospinprop}), but $\underline{\bsy{\alpha}}$ is now the vector of the \textit{classical} reflection points.
The $g_N^\text{sc}(\bsy{x}, \bsy{x'})$ are well known, see e.g. Refs. \onlinecite{Gutzwiller1990} and \onlinecite{Baranger1993a}.
The stationary phase condition selects all sets of $N$ stationary boundary points minimizing the phase aquired, and hence specifies \textit{classical trajectories} of the system. We thus obtain our final expression for $G^\text{sc}(\bsy{x},\bsy{x'})$ in terms of a sum over classical trajectories $\gamma$ that connect the
points $\bsy{x'}$ and $\bsy{x}$:
\begin{equation}
\label{eq:Gsc1}
 \Gsc(\bsy{x},\bsy{x'}) = \frac{\hbar \vF}{2}\!\!\sum_{\gamma(\bsy{x},\bsy{x'})} \!\frac{|D_{\gamma}|}{\sqrt{2\pi\hbar^3}}
e^{i\kE L_{\gamma}+i\mu_{\gamma}\pi/2}\,K_{\gamma}\,.
\end{equation}
Here, $L_{\gamma}$, $\mu_{\gamma}$ and $N_{\gamma}$ are the length, the number of conjugate points and the number of reflections at the boundary for the
classical orbit $\gamma$. $K_\gamma= K_{N_\gamma}$ is the corresponding pseudospin propagator and  
\begin{equation}
 D_\gamma =
\frac{1}{\vF}\left|\left(\frac{\partial{x}_{\perp}}{\partial {p'}_\perp}\right)\right|^{-1/2}_{\gamma}\,.
\end{equation}
measures the stability
of the path $\gamma$ starting at $\bsy{x'}$ with momentum $\bsy{p'}$ and ending at $\bsy{x}$ with momentum $\bsy{p}$.
The $\perp$ denotes that the derivative involves only the projections perpendicular to the trajectory, which are scalars in two dimensions.

 Expression (\ref{eq:Gsc1}) represents one main result of the present paper: The semiclassical charge dynamics for electrons and holes in a ballistic graphene flake is very similar to the case of electrons in Schr\"odinger billiards with Dirichlet boundary conditions. The graphene specific physics is incorporated in the pseudospin dynamics described by $K_\gamma$. 

For a trajectory containing only one single reflection we have
\begin{eqnarray}
\tilde{K}_{\gamma}^{(1)}  &=& (1+\sigma_{\bsy{x\alpha}})\sigma_{\bsy{n_{\alpha}}}
\mathcal{R}_{\bsy{\alpha}} P_{\bsy{\alpha}}
(1+\sigma_{\bsy{\alpha x'}})\,.
\end{eqnarray}
Using the classical relations between the vectors $\bsy{x}- \bsy{\alpha}$ and $\bsy{\alpha}-\bsy{x}'$ yields
\begin{eqnarray}
\label{eq:K^1}
 K_{\gamma}^{(1)}  &=& \pm i\, \bsy{\nu}\cdot\bsy{\tau} \nonumber \\
&\otimes&
\left\{
\begin{array}{cl}
 e^{\pm i\theta \tau_z}\sigmav{\bsy{t_\alpha}} (1+\sigma_{\bsy{\alpha x'}}) & \text{for~zz}\,, \\
 e^{i\theta \sigma_z}\sigma_z (1+\sigma_{\bsy{\alpha x'}})& \text{for~ac~and~im}\,.
\end{array}
\right.
\end{eqnarray}
with $\bsy{\nu}$ according to Tab.\,\ref{tab:bc}.
With this result, we can obtain the pseudospin propagator for an arbitrary number of reflections by iteration.


\subsection{Trace formulae and semiclassical shell effects for classically integrable graphene billiards}
\label{ssec:regular}

In this section we give two representative examples for trace formulae describing the oscillating part of the 
density of states in graphene billiards that have classically integrable dynamics: circular and rectangular billiards with different types of graphene edges. We derive the corresponding semiclassical trace formula for the class of classically chaotic graphene cavities in Sec. \ref{ssec:GTF}. 

\begin{figure}
 \centering
 \includegraphics[width=0.18\textwidth]{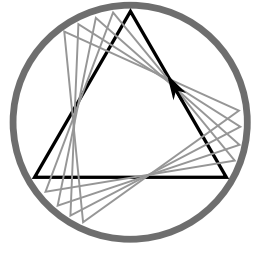}
\caption{\small{Example of a family of degenerate classical orbits in a circular billiard. The black triangular orbit can be rotated by an arbitrary angle
without changing its length. All resulting orbits contribute the same to the density of states. (adapted from Ref. \onlinecite{Brack2008}.)}}
\label{Fig:DegCircle}
\end{figure}

Orbits in regular systems are organized in families on classical invariant tori. An example of such a (periodic) orbit family is sketched for the circular billiard in Fig.\,\ref{Fig:DegCircle}. The members of a family possess the same classical properties entering Eq.\,(\ref{eq:Gsc1}) such as action, length, stability, number of reflections and number of conjugate points. In order to compute the oscillatory part of the DOS from the semiclassical Green's function it is convenient to organize the trajectories in terms of tori, respectively families $f$, in the trace-integral, Eq.\,(\ref{eq:dos2}):
\begin{equation}
   \rho(\kE)  = -\frac{1}{\pi}\mathfrak{Im} \sum_f \intl_{\mathcal{V}_f} \!d\bsy{x} \,\text{Tr}\left[G_f(\bsy{x},\bsy{x})\right]
\end{equation}
leading to the Berry-Tabor formula for $\rho_\text{osc}$ in terms of sums over families of periodic orbits organized on resonant tori\cite{Berry1976/77}.
The semiclassical pseudospin propagator for graphene does not alter the resonance condition (cf. the chaotic case \ref{ssec:GTF}) , and for periodic classical orbits its trace $\tr\left(K_\gamma\right)$ does not depend on the coordinates of the starting and end point:
\begin{eqnarray}
\label{eq:pspin_periodic}
 \sigmav{\bsy{\alpha}_1 \bsy{x}} =  \sigmav{\bsy{x}\bsy{\alpha}_N} =  \sigmav{\bsy{\alpha}_1 \bsy{\alpha}_N}\,.
\end{eqnarray}
 Therefore, the integrals over $\mathcal{V}_f$ are the same as for Schr\"odinger billiards with Dirichlet boundary conditions. Hence we can adapt the corresponding results by explicitly including the correct pseudospin trace for each orbit family.

The collective effect of orbit families giving rise to constructive interference due to action degeneracies lead to pronounced signatures in the DOS of integrable systems known as shell effects\cite{Brack2008}. We analyze below how such features are modified due to graphene edge effects.
\begin{table}
 \begin{tabular}{c|c|c||c|c||c|c}
\multicolumn{3}{c||}{a) circular infinite} & \multicolumn{2}{c||}{ b) square billiard} & \multicolumn{2}{c}{c) square billiard} \\ 
\multicolumn{3}{c||}{mass billiard} &  \multicolumn{2}{c||}{(``semiconducting'')} & \multicolumn{2}{c}{ (``metallic'') } \\ 
\hline
\hspace*{0.05cm}TF \hspace*{0.05cm}  & TF (P) 
 &  \hspace*{0.05cm} QM \hspace*{0.05cm} & \hspace*{0.3cm}TF \hspace*{0.3cm} & QM & \hspace*{0.2cm}TF \hspace*{0.2cm} & QM \\
\hline
1.49 & 1.57  & 1.43    & 6.85  & 6.81    & 6.86  & 6.85    \\
2.72 & 2.78  & 2.63    & 7.85  & 7.84    & 7.30  & 7.28    \\
3.10 & 3.14  & 3.11    & 7.93  & 7.87    & 7.92  & 7.85    \\
3.87 & 3.92  & 3.77    & 8.11  & 8.05    & 8.15  & 8.09    \\
4.46 & 4.49  & 4.48    & 8.97  & 8.92    & 8.41  & 8.39    \\
4.69 & 4.71  & 4.68    & 9.11  & 9.10    & 8.84  & 8.80    \\
5.00 & 5.04  & 4.88    & 9.26  & 9.24    & 9.43  & -       \\
5.73 & 5.75  & 5.75    & 9.35  & 9.32    & 9.54  & 9.50    \\
6.10 & 6.12  & 5.98    & 10.47 & -       & 9.85  & 9.85    \\
6.10 & 6.14  & 6.09    & 10.86 & 10.86   & 10.06 & 10.05   \\
6.26 & 6.28  & 6.27    & 10.92 & 10.90   & 10.59 & 10.56   \\
6.95 & 6.98  & 6.98    & 11.05 & 11.01   & 11.04 & 11.00   \\
7.20 & 7.23  & 7.06    & 11.18 & 11.14   & 11.04 & 11.03   \\
7.43 & 7.45  & 7.41    & 11.29 & 11.27   & 11.21 & 11.16   \\
7.71 & 7.72  & 7.71    & 11.52 & -       & 11.71 & 11.69   \\
\end{tabular}

\caption{a) Energy levels $k_n R$ of the circular billiard with infinite mass type edges obtained from the semiclassical trace formula Eq.\,(\ref{eq:TF_circle}) by summing over many classical orbits with $\xi=0$ (TF) and by summing up all orbits approximately (TF (P)) Eq.\,(\ref{eq:circle_poisson}) compared to the quantum mechanical result (QM) Eq.\,(\ref{eq:EV_circle_im}).
b), c) Energy levels $k_n L$ for square billiards with $K L \mod 2\pi= 2\pi/3$ ($L=200\,a$ ``semiconducting'') and $K L \mod 2\pi = 0$ ($L=201\,a$ ``metallic''), respectively. Again we compare the result from the semiclassical trace formula (\ref{eq:TF_Rect_SC}) at $\xi=0$ with the quantum mechanical result (\ref{eq:quantiz_rect}).}
 \label{tab:peaks}
\end{table}
\subsubsection{Circular billiard with infinite mass type edges}
We begin with a circular billiard with infinite mass type edges. Then the quantum energy levels $E_{nm} = \hbar \vF k_{nm}$ are given by 
the intersections of Bessel functions\cite{Berry1985}
\begin{equation}
\label{eq:EV_circle_im}
 J_n(k_{nm} R) = \tau J_{n+1}(k_{nm} R)\,,
\end{equation}
where $R$ is the billiard radius, $\tau = \pm 1$ labels the two valleys and $n,m\in \mathbb{Z}$, where $m$ counts the intersections.

\begin{figure}
 \centering
 \includegraphics[width=0.35\textwidth]{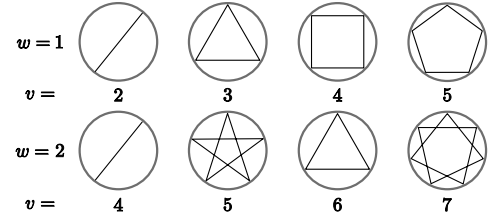}
\caption{\small{Classical periodic orbits representing families in the circular billiard. $v$ is the total number of reflections along the orbit and $w$ denotes the winding number. If ($v,w$) are not coprime the orbit is a repetition of a shorter primitive orbit. E.\,g. (4,2) is a repetition of (2,1) and (6,2) of (3,1).
(Adapted from Ref. \onlinecite{Balian1972}.) }}
\label{Fig:OrbitsCircle}
\end{figure}

For the semiclassical calculation of $\rho_\text{osc}$ we adapt results for the Schr\"odinger disk billiard as derived and discussed in detail e.g. in Ref. \onlinecite{Brack2008}. Periodic orbit families in the disk are labeled by the total number of reflections $v$ and the winding number $w$, with $v\geq 2w$. Examples with $w=1,2$ are depicted in Fig.\,\ref{Fig:OrbitsCircle}.
We also allow for negative winding numbers $w$, and define the sign such that $w>0$ for clockwise going orbits and $w<0$ for anti-clockwise going orbits. Simple geometry gives for the length $L_{v,w}$ and the angle of rotation $\varphi_{v,w}$ aquired of an orbit $(v,w)$ 
\begin{eqnarray}
\label{eq:circle_poisson}
 L_{v,w} &=& 2 v R \sin(|\varphi_{v,w}|)\,, \\
\varphi_{v,w} &=& \pi\frac{w}{v}\,.
\end{eqnarray}
Then the reflection angles read
\begin{equation}
 \theta_{v,w} = \left(\frac{\text{sgn}(w)}{2} - \frac{w}{v}\right) \pi\,.
\end{equation}
Graphene physics enters through the pseudospin propagator, Eq.\,(\ref{eq:Kgamma}), with boundary matrix 
\begin{equation}
 P_{\bsy{\alpha}}=(1+\tau_z \otimes\sigmav{\bsy{t_\alpha}})/2
\end{equation}
for the infinite mass case [see Eq.\,(\ref{eq:bc}) and Tab.\,\ref{tab:bc} \,]. For an orbit $(v,w)$
the trace over $K$ yields
\begin{eqnarray}
\label{eq:trace_im_circ}
 \tr{K}_{v,w} &=& i^v\tr\left(\tau_z^v \otimes \sigma_z^v e^{iv \theta_{v,w}\sigma_z} \right) \nonumber \\
&=& 4 \cos(v\,\theta_{v,w})
\left\{
\begin{array}{cl}
 (-1)^{v/2} & \text{for~even~}v\,, \\
 0 & \text{for~odd~}v\,.
\end{array}
\right.
\end{eqnarray}
Equation (\ref{eq:trace_im_circ}) reveals the interesting property
that only orbits with an even number of reflections are contributing to the oscillating DOS in the circular graphene billiard, while
for odd $v$, the pseudospins are interfering destructively. Note that this holds true also in each valley separately, because in the case of odd $v$, the contributions from winding numbers $w$ and $-w$ have opposite signs.

Adapting the expression for the circular Schr\"odinger billiard\cite{Brack2008, Bogachek1973} accordingly yields the semiclassical expression for the oscillatory part of the DOS of the graphene disk:
\begin{eqnarray}
\label{eq:TF_circle}
 \rho_{\text{osc}}^{\text{sc}}(\kE) &=&
4\sqrt{\frac{\kE R^3}{\pi}}  \sum_{w=1}^{\infty}\sum_{\stackrel{\scriptstyle v=2w}{\text{even}}}
^{\infty} (-1)^{w+v/2}\frac{f_{v,w}}{\sqrt{v}}  \\ &&\times
\sin^{3/2}(\varphi_{v,w})
\sin\left(\kE L_{v,w} + \frac{3}{4}\pi\right) e^{-(\xi L_{v,w} /2 )^2} \nonumber
\end{eqnarray} 
where $f_{v,w}=1$ if $v=2w$ and otherwise  $f_{v,w}=2$.

The last factor in Eq.\,(\ref{eq:TF_circle}), giving rise to an exponential suppression of orbits of length $L_{v,w} > 1/\xi$, represents a broadening of the peaks in the quantum density of states by convoluting $\rho$ with a Gaussian of width $\xi$. Such a broadening is additionally introduced to mimic e.g. temperature smearing or account for a finite life time of the quantum states, for instance due to residual disorder scattering\cite{Richter1996}. Thereby, Eq.\,(\ref{eq:TF_circle}) relates gross effects in smeared quantum spectra or experimental spectra obtained with limited resolution to the contributions from families of shortest periodic orbits \cite{Brack2008,Richter1996a}.

Using the Poisson summation formula, we can approximately sum up the trace formula (\ref{eq:TF_circle}) for $\xi=0$ and find the approximate eigenenergies $k_{VW} = x_{VW}/R$ corresponding to poles in the semiclassical sum, that fulfill the equation 
\begin{eqnarray}
V + \frac{3}{2}   &=& (2W+1)[1-\arccos(W/X_{VW})/\pi] \nonumber \\ && + \frac{2X_{VW}}{\pi} \sqrt{1-W^2/X_{VW}^2} - 2W \,.
\end{eqnarray}

\begin{figure}
 \centering
 \includegraphics[width=0.45\textwidth]{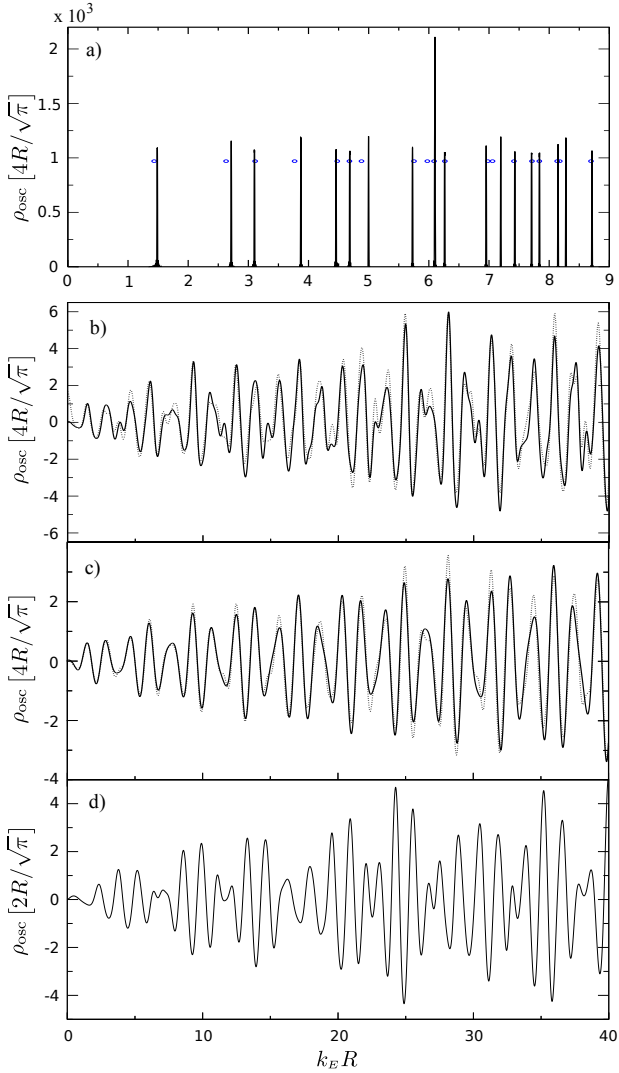}
\caption{\small{(color online). Oscillating part $\rho_\text{osc}$ of the density of states of a circular billiard as a function of $\kE R$. 
a) Peaks are obtained from the semiclassical expression (\ref{eq:TF_circle}) by summing up orbit families up to $v, w = 400$ for $\xi=0$. Blue circles mark the positions of the exact quantum mechanical levels given by Eq.\,(\ref{eq:EV_circle_im}) (See also Tab\,\ref{tab:peaks}\,a)).
b) Gaussian convoluted $\rho_\text{osc}$ for $\xi= 0.3/R$. The full (dotted) curves show the semiclassical (quantum mechanical) results. 
c) Comparison between the full semiclassical orbit sum (dotted, $\xi= 0.4/R$) with the contribution from the two shortest orbit families $(2,1)$ and $(4,1)$ (solid).
d) Corresponding results (for $\xi= 0.4/R$) for a circular Schr\"odinger billiard with Dirichlet boundary conditions.
}
}
\label{Fig:ODOS_Circle}
\end{figure}

In Fig.\,\ref{Fig:ODOS_Circle}\,a)-c) we compare the results of the semiclassical trace formula (\ref{eq:TF_circle}) with exact quantum results from Eq.\,(\ref{eq:EV_circle_im}) for the lower part of the graphene disk spectrum. For $\xi=0$ [panel a)] even the exact quantum levels (blue circles) are reproduced with remarkable accuracy by the semiclassical theory [black peaks, see also numerical values in Tab\,\ref{tab:peaks}\,a)]. For every level, we have a sharp peak in the semiclassical result. An exception are the two levels
close to $\kE R = 6$, for which we have only one peak, though twice as high as the others, meaning that in the semiclassical expression the
two levels are nearly degenerate. 

Panel b) shows the broadened spectrum for $\xi=0.3/R$. Again, the semiclassical result (solid line) is in very good agreement with the corresponding quantum result (dotted). For comparison, panel d) shows the same energy range for the corresponding Schr\"odinger billiard. In Fig.\,\ref{Fig:ODOS_Circle}\,c) we have a closer look at which orbit families contribute. In fact we can see from Fig.\,\ref{Fig:ODOS_Circle}\,c) that the two shortest non-vanishing orbit families $(2,1)$ and $(4,1)$ already yield a good approximation to the shell structure for $\xi=0.4/R$.

\begin{figure}
 \centering
 \includegraphics[width=0.45\textwidth]{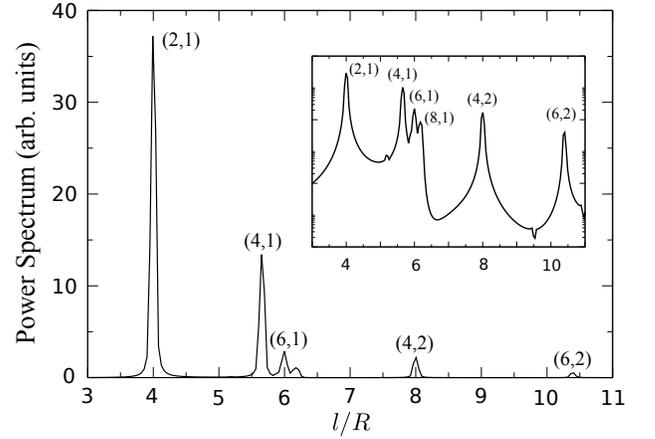}
\caption{\small{Power spectrum of the Gaussian convoluted \mbox{($\xi=0.3/R$)} quantum density of states of the graphene disk with infinite mass edges. Peaks can be uniquely assigned to periodic orbit families $(v,w)$, see text. Inset: Logarithmic respresentation.
}}
\label{Fig:ODOS_Circle_FT}
\end{figure}

 Fig.\,\ref{Fig:ODOS_Circle_FT} shows the power spectrum of the exact quantum result (Gaussian convoluted with $\xi=0.4/R$). Evidently, only families with an even number of vertices $v$ are contained in the spectrum, as semiclassically predicted. For example the triangular orbits $(3,1)$ that would give a peak at $L/R=5.2$ and also the pentagram orbits $(5,2)$ ($L/R=9.5$) do not contribute. The inset shows the same plot on a logarithmic scale, where the absence of the odd orbits is even more evident.
\subsubsection{Rectangular billiard with zigzag and armchair edges}
\begin{figure}
 \centering
 \includegraphics[width=0.48\textwidth]{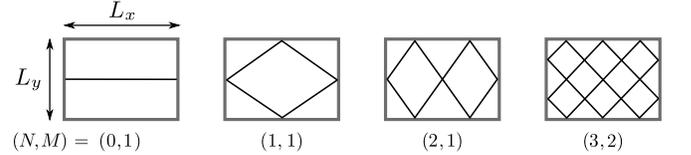}
\caption{\small{Families of periodic classical orbits in the rectangular billiard. $N (M)$ is the number of reflections at the bottom (left) side.}}
\label{Fig:OrbitsRectangle}
\end{figure}
The rectangular billiard represents another prominent classically integrable geometry. While for the Schr\"odinger equation with Dirichlet boundary conditions this is a simple textbook problem, there is no explicit expression for eigenenergies of the graphene rectangle with two opposite zigzag and two opposite armchair edges. (For the derivation of a closed formula for the quantum eigenenergies in terms of a transcendental equation see App.\,\ref{app:Rectangle_ev}). We will show that our semiclassical theory provides a very good approximation to the quantum density of states.

In the rectangle, the periodic orbit families can again be labeled with two indices. We denote by $N$ and $M$ the number of reflections at the bottom zigzag ($N$) and the left armchair ($M$) side of the rectangle with lengths $L_x$ and $L_y$ respectively (see Fig.\,\ref{Fig:OrbitsRectangle}).
The absolute values of the reflection angles at the zigzag and armchair edges then read
\begin{equation}
\begin{split}
&  |\theta_{\text{zz}}| = \arctan\left(\frac{M L_x}{N L_y}\right)\,, \\ & |\theta_{\text{ac}}|
 = \frac{\pi}{2} - |\theta_{\text{zz}}| = \arctan\left(\frac{N L_y}{M L_x}\right) \,.
\end{split}
\end{equation}
From Eq.\,(\ref{eq:K^1}) we can read off the following matrix factors for reflections with angles $\theta_{\text{zz}}$ and $\theta_{\text{ac}}$, respectively:
\begin{equation}
\begin{split}
 -i\tau_z e^{-i\theta_{\text{zz}} \tau_z} \otimes \sigma_x &~~~~ \text{lower zigzag edge,} \\
 -i\tau_z e^{i\theta_{\text{zz}} \tau_z} \otimes \sigma_x &~~~~ \text{upper zigzag edge,} \\
i\tau_y \otimes \sigma_z e^{i\theta_{\text{ac}}\sigma_z} &~~~~ \text{left armchair edge,} \\
-i\tau_y e^{i2KL_x \tau_z} \otimes \sigma_z e^{i\theta_{\text{ac}}\sigma_z} &~~~~ \text{right armchair edge}\,.
\end{split}
\end{equation}
This enables us to calculate the pseudospin trace of a periodic orbit from family $(N,M)$
as
\begin{equation}
\label{eq:Trace_Rect}
 \tr {K}_{NM} = (-1)^N 4 \cos(2MKL_x - 2N|\theta_{\text{zz}}|)\,.
\end{equation}
This expression holds irrespective of the propagation direction along the orbit.
Note also that the $\theta_{\text{zz}}$ in Eq.\,(\ref{eq:Trace_Rect}) occurs only due to the fact that we have different zigzag edges at the top and the bottom boundary (A- and B-terminated, respectively).
Equation (\ref{eq:Trace_Rect}) is now used to adapt the trace formula for the Schr\"odinger equation which has been derived e.\,g. in Refs. \onlinecite{Brack2008} and
\onlinecite{Richter1996a} to the case of graphene. Taking into account the interfering pseudospins in graphene, we find
\begin{eqnarray}
\label{eq:TF_Rect_SC}
 \rho_{\text{osc}}^{\text{sc}}(\kE) &=& \sqrt{\frac{\kE}{2\pi^3}} \sum_{M=1}^{\infty} \sum_{N=1}^{\infty}
\frac{ f_{NM}\, L_xL_y }{\sqrt{L_{NM}}}  \\ &&\times
\cos\left(\kE L_{NM} - \frac{\pi}{4}\right) \tr {K}_{NM} \, e^{-(\xi L_{NM} /2 )^2} \nonumber
\end{eqnarray}
with length $L_{NM} = 2 \sqrt{M^2L_x^2+N^2L_y^2}$ and $\tr {K}_{NM}$ from Eq.\,(\ref{eq:Trace_Rect}). Further $f_{NM}=1$ if $N=0$ or $M=0$ and otherwise $f_{NM}=2$.
Note that the size of the billiard determines whether certain orbits contribute: The quantity $K L_x$ can only take values that are multiples of $\pi/3$. In particular for $K L_x = 0 \mod 2\pi$ \cite{note_3}, families $(N,N L_y/L_x)$ with odd $N$ do not contribute according to Eq.\,(\ref{eq:Trace_Rect}).
Further examples are the families $(M,0)$ and $(0,N)$ for odd $N$ and $M$ respectively.
They cancel each other exactly for $K L_x = 0 \mod 2\pi$ because of the $(-1)^N$ term in the pseudospin trace.

\begin{figure*}
 \centering
 \includegraphics[width=0.9\textwidth]{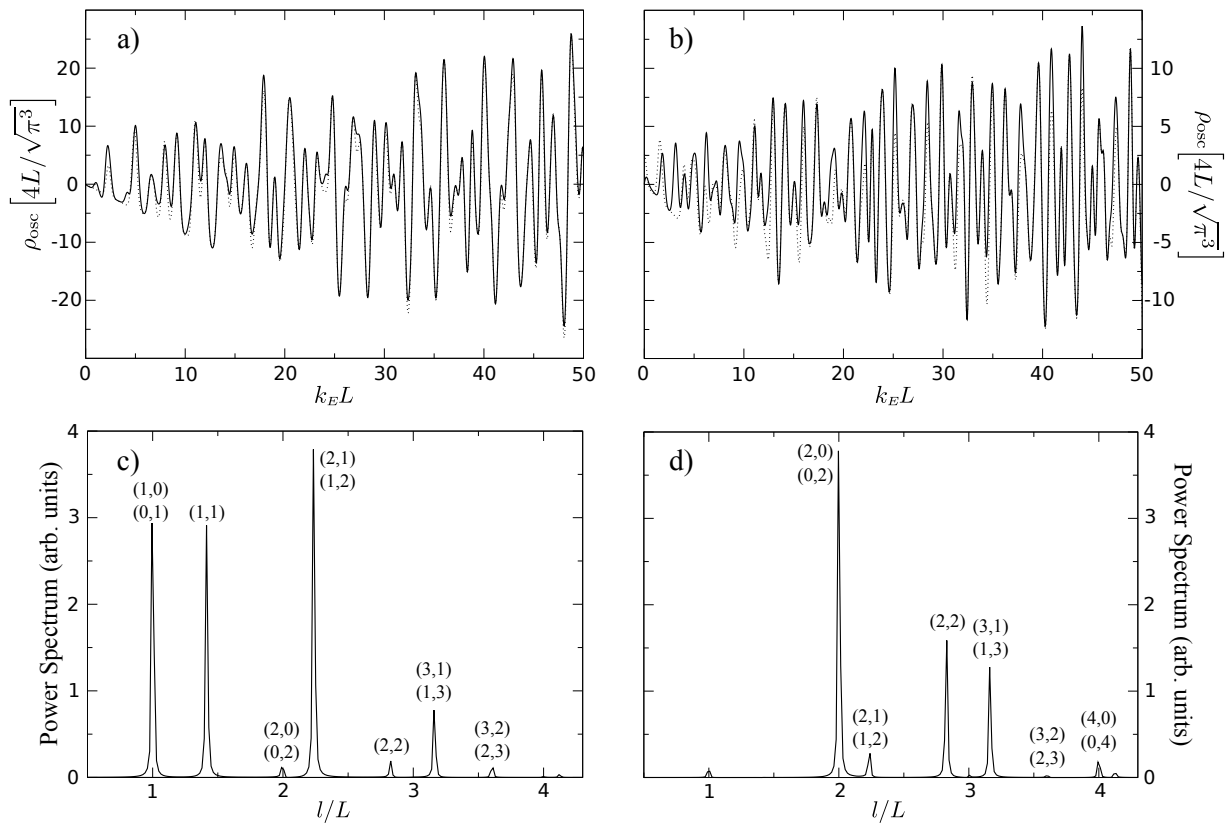}
\caption{\small{Oscillating part of the density of states of a square billiard with two armchair edges and two zigzag edges ($L_x=L_y=L$).
The left panels (a, c) show the results for a square with $K L \mod 2\pi = 2\pi/3$ (``semiconducting'') and the right panels (b, d)
are for a square with $K L \mod 2\pi = 0$ (``metallic'').
Panels a) and b) show the Gaussian convoluted $\rho_{\text{osc}}(\kE)$
for $\xi=0.3/L$. The dotted curves represent the quantum mechanically exact results
calculated with Eq.\,(\ref{eq:quantiz_rect}) and broadened correspondingly. 
Panels c) and d) show the quantum mechanical power spectra (for $\xi=0.3$).
It is easy to identify the peaks associated with corresponding families $(M,N)$.
}}
\label{Fig:ODOS_Square}
\end{figure*}

In Fig.\,\ref{Fig:ODOS_Square} and Tab\,\ref{tab:peaks}\,b),\,c) we compare the results from the semiclassical trace formula (\ref{eq:TF_Rect_SC}) for $L_x=L_y=L$ 
with the quantum mechanical results obtained by solving Eq.\,(\ref{eq:quantiz_rect}) numerically.
Again we find very good agreement with the quantum result. This is rather remarkable because of the complicated structure of the quantization condition (\ref{eq:quantiz_rect}). The semiclassical predictions concerning the frequency content of the DOS oscillations are confirmed in
Fig.\,\ref{Fig:ODOS_Square}\, c) and d). For example the shortest orbits $(1,0)$, $ (0,1)$ and $(1,1)$ do not contribute for the system in d) ($K L \mod 2\pi = 0$) due to destructive pseudospin interference, while they are important in c) ($K L \mod 2\pi = 2\pi/3$).

Note that in Tab.\,\ref{tab:peaks} we find some additional levels from the semiclassical trace formula, which cannot be associated to quantum energy levels of the rectangle. Rather these peaks occur at positions that fulfill the quantization condition of a fictitious 1D quantum well of width $L$ with armchair boundary conditions. It is well known\cite{Brack2008} that this is an effect of subleading order ($[\kE L]^{-1/2}$ with respect to leading order) produced by orbits that \textquoteleft graze\textquoteright ~along the edges.
\subsection{Trace formula for classically chaotic graphene billiards}
\label{ssec:GTF}
Finally we consider classically chaotic graphene systems. In this case no spatial symmetries are present that would give rise to an orbit degeneracy as in the regular case. From Eq.\,(\ref{eq:Gsc1}) we already know that the final result differs from the trace formula for chaotic Schr\"odinger billiards only with respect to the pseudospin trace. Thus we have to work out how the spatial integral in Eq.\,(\ref{eq:dos2}) depends on this trace. To this end we do not start directly from the semiclassical Green's function (\ref{eq:GN_2}), but go one step back to Eq.\,(\ref{eq:GN_1}).
In order to calculate the integral
\begin{equation}
\label{eq:rho_IG}
   \rho_N(\kE)  = -\frac{1}{\pi}\mathfrak{Im} \intl_{\mathcal{V}} \!d\bsy{x} \,\text{Tr}\left[G_N(\bsy{x},\bsy{x})\right]
\end{equation}
we consider only the $\bsy{x}$-dependent part of the integrand,
\begin{equation}
\label{eq:traceint}
 I_N =  \intl_{\mathcal{V}} \!d\bsy{x} \,\frac{K(\bsy{x},\underline{\bsy{\alpha}})}{\sqrt{|\bsy{x}-\bsy{\alpha}_N||\bsy{\alpha}_1-\bsy{x}|}}
 e^{i\kE(|\bsy{x}-\bsy{\alpha}_N|+|\bsy{\alpha}_1-\bsy{x}|)}\,,
\end{equation}
 and choose the parametrization $\bsy{x} = l\,\hat{\bsy{l}}+t\,\hat{\bsy{t}}$, where $\hat{\bsy{l}}$ is the direction from $\bsy{\alpha}_N$ to $\bsy{\alpha}_1$
 and $\hat{\bsy{t}}$ the direction perpendicular to $\hat{\bsy{l}}$ such that a right handed coordinate system results. The origin $l=t=0$ is at the point $\bsy{\alpha}_N$
 and we denote $l_{N1} = |\bsy{\alpha}_N -\bsy{\alpha}_1|$. Then we can rewrite the phase
\begin{eqnarray}
\varphi(l,t)/\kE \!\!\!\!\! &=& \!\!\!\!\!  |\bsy{x}-\bsy{\alpha}_N|+|\bsy{\alpha}_1-\bsy{x}| \nonumber \\
&\stackrel{t\ll l, l-l_{N1}}{\approx}& l_{N1} \left(1+\frac{t^2}{2l[l_{N1}-l]}\right)\,.
\end{eqnarray}
We are now evaluating the $t$-integral in stationary phase approximation assuming $\kE l_{N1} \gg 1$. The stationary phase point $t_0$ is given by
\begin{eqnarray}
 \frac{\partial \varphi(l,t_0)}{\partial t} &=& \frac{\kE \,l_{N1}t_0}{l(l_{N1}-l)} = 0 \quad \Rightarrow \quad t_0 = 0\,, \\
 \frac{\partial^2 \varphi(l,t_0)}{\partial t^2} &=& \frac{\kE\, l_{N1}}{l(l_{N1}-l)}\,, \\
\quad \varphi(l,t_0) &=& \kE\, l_{N1} = \kE \,|\bsy{\alpha}_N-\bsy{\alpha}_1|\,.
\end{eqnarray}
This means however that at the critical point $t_0$, the pseudospin propagator $K(\underline{\bsy{\alpha}})$ has no dependence on $l$ left, since for $t=0$ Eq.\,(\ref{eq:pspin_periodic}) holds.
Thus the remaining integral can be performed exactly:
\begin{eqnarray}
 I_N 
=\sqrt{\frac{2\pi l_{N1}}{\kE}} K(\underline{\bsy{\alpha}})\, e^{i\kE \,|\bsy{\alpha}_N-\bsy{\alpha}_1|}\,.
\end{eqnarray}
This tells us that as for the Green's function, we can essentially read off the result for $\rhoosc^{\text{sc}}$ directly from the
corresponding Dirichlet problem for the Schr\"odinger equation\cite{Gutzwiller1990} and find the Gutzwiller-type trace formula for a chaotic graphene cavity
\begin{equation}
\label{eq:Gutzwiller}
  \rho_{\text{osc}}^{\text{sc}}(\kE) = \frac{\vF}{2\pi}\mathfrak{Re}\sum_{\gamma}
 \tr( K_\gamma) A_\gamma \,e^{i\kE L_\gamma}\,.
\end{equation}
Here the sum runs over all, infinitely many classical periodic orbits $\gamma$, because the stationary phase points with $t=t_0=0$ are lying exactly on the straight line
connecting the last with the first reflection point, i.\,e. the apperance of the pseudospin does not affect the stationary points. The classical amplitudes $A_\gamma$ depend on the period, the stability and the Maslov index of the corresponding orbit\cite{Gutzwiller1990}.
That means, except for $\hbar$ and the trace over $K_\gamma$, accounting for the interference of pseudospins, the right-hand side of Eq.\,(\ref{eq:Gutzwiller}) contains only classical quantities and has the same structure as Gutzwiller's trace formula. We note that in Ref. \onlinecite{Carmier2008} a semiclassical trace formula is presented for $\rho_{\text{osc}}$, which however is not taking into account the boundaries required to obtain chaotic dynamics. Note that the expression (\ref{eq:Gutzwiller}) for $\rho_{\text{osc}}$ is only valid for systems with isolated orbits, a prerequisite to evaluate the integral perpendicular to $\bsy{\alpha}_N- \bsy{\alpha}_1$ in stationary phase approximation. This is particularly fulfilled for chaotic systems. 

Expression (\ref{eq:Gutzwiller}) allows, in principle, for computing semiclassical approximations for energy levels in chaotic graphene billiards. We presume that this trace formula holds true more generally for classically chaotic graphene systems, not only billiards, with an appropriate generalization of the pseudospin evolution.
Since the classical dynamics of a graphene billiard is the same as that of a Schr\"odinger billiard, the convergence properties of Eq.\,(\ref{eq:Gutzwiller}) are expected to be similar to those of Gutzwiller's trace formula, with convergence problems linked to the exponential proliferation of periodic orbits with their length. 
In App. \ref{app:BulkDis}, we discuss the effect of weak bulk disorder on the trace formula (\ref{eq:Gutzwiller}).

As Gutzwiller's trace formula for the case of quantum chaotic Schr\"odinger dynamics, the trace formula (\ref{eq:Gutzwiller}) represents a suitable starting point to consider the statistical properties of energy levels for chaotic graphene cavities, in particular universal spectral features within certain symmetry classes. Based on Eq.\,(\ref{eq:Gutzwiller}) we devote a major part of Ref. \cite{partII} to the semiclassical analysis of spectral statstics in graphene. There we will see that intervalley scattering, semiclassically incorporated in the pseudospin dynamics, plays a key role for the effective symmetry class obeyed in graphene, e.\,g. unitary, orthogonal or intermediate statistics between the two.

\section{Conclusion}
\label{sec:conclusion}

The growing ability to manufacture graphene-based nanostructures
and their increasing role in the field of graphene physics poses
challenges to theory to treat confinement effects. Addressing ballistic
graphene cavities we have focussed on the effect of different types
of edges, zigzag, armchair and inifinite mass type, on the spectral properties. 
The multiple reflection expansion used, combined with the semiclassical 
approximation, allows for incorporating and analyzing edge phenomena in 
a particularly transparent way, both for the mean density of states $\bar{\rho}$ 
as well as for the remaining oscillatory part: The leading-order
Weyl contribution to $\bar{\rho}$ for graphene billiards scales with the phase 
space volume on the energy shell, as for Schr\"odinger-type billiards.
Edge effects are expected to alter the perimeter correction to  
$\bar{\rho}$, which is proportional to the total boundary length
in the Schr\"odinger case with Dirichlet boundary conditions.
We showed for graphene billiards that armchair and infinite mass edges
do not give any perimeter contribution, while zigzag edges yield a 
characteristic low-energy term scaling with the length of the zigzag boundary.
As analyzed in detail we could  relate this boundary
term in $\bar{\rho}$ to the average number of quantum zigzag edge states.
Thereby, our approach allows for an alternative, analytical calculation
of the zigzag edge state contribution. For graphene nanostructures 
with unknown portion of zigzag-type edge segments, this enables one to
estimate the effective zigzag edge length, respectively number of edge states,
from the characteristic feature in $\bar{\rho}(E)$, see Figs. \ref{Fig:WeylNNN} and \ref{Fig:WeylNN}.
Hence, already the mean density of states of graphene flakes incorporates important physical 
information.

For the oscillatory contribution, $\rho_{\rm osc}$, to the density of states
of graphene billiards we derive semiclassical trace formulae in terms of sums
over classical periodic orbits. We show that, within the leading-order
semiclassical approximation, the classical orbital dynamics entering into
the semiclassical sums is the same as for Schr\"odinger billiards of the same geometry. 
This implies for regular graphene geometries Berry-Tabor like \cite{Berry1976/77} sums
over families of orbits and for chaotic geometries a Gutzwiller type \cite{Gutzwiller1990}
trace formula in terms of isolated periodic trajectories. Edge effects
enter into the contribution of each periodic orbit (family) exclusively through the
the pseudospin propagator and its trace along the orbit. This leads to
a particularly transparent representation of graphene edge phenomena. We gave a detailed
interpretation for two representative regular systems: the graphene disk with 
infinite mass edges and the graphene 2d box with boundaries built from two
zigzag and two armchair edges. The comparison with full quantum results showed
very good agreement, both for smeared spectra, highlighing the role
of short, fundamental periodic orbits, and on the level of individual energy levels,
obtained semiclassically by summing up many orbit families.

A number of questions and further research directions is now arising from this work. 
They include the challenge to generalize the semiclassical expressions for
the density of states of clean billiards to cavities with impurity scattering
and systems with smooth confinement potentials, more generally graphene
with arbitrary classical Hamiltonian dynamics, including also systems with mixed
phase space. 
Second, the fact that our treatment of the zigzag edge associated
average level density proofs adequate for both settings, models without and with
particle-hole breaking effects, e.g. from next-nearest-neighbor coupling, see Sec. \ref{sssec:Weylzz},
encourages to address zigzag edge magnetism \cite{Fujita1996, Wimmer2008, Son2006, Tao2011} within this framework.
Third, the semiclassical formalism
developed allows for treating graphene nanostructures with boundaries 
that can be viewed of being composed of many zigzag- and armchair-edge segments.
In particular, analytical expressions can be derived by treating long orbits with
bounces off the different boundary segments in a statistical way.
Fourth, the techniques used can be generalized to quantum transport through
open graphene nanostructures. 

In a second paper \cite{partII} we will particularly
address the two last items and study spectral statistics (through the spectral
form factor) of closed systems and transport properties (weak localization,
universal conductance fluctuations and shot noise) of open graphene billiards.

\section{Acknowledgements}

We thank Philippe Jacquod, Viktor Kr\"uckl, Jack Kuipers, Juan Diego Urbina and Michael Wimmer
for useful conversations.
We acknowledge funding through the Deutsche For\-schungsgemeinschaft
within DFG Research Training Group 1570 (KR, JW) and through TUBA under 
grant I.A/TUBA-GEBIP/2010-1  and the funds of the Erdal \.{I}n\"on\"u chair at Sabanc\i~University (IA).
JW further acknowledges the support and hospitality at Sabanc\i~University.\\

\appendix

\section{The singularity of a Dirac-charge layer}
\label{app:jump}
Here we derive the expression (\ref{eq:jump}) inducting the discontinuity of the Green's function at the boundary\cite{Adagideli2002}. Using the short distance asymptotic form for the Hankel function
\begin{equation}
 H_0^{+}(\xi) \stackrel{\xi\ll 1}{\longrightarrow} \frac{2i}{\pi}\ln(\xi/2)\,,
\end{equation}
we obtain the short range singularities of the free Green's function from Eq.\,(\ref{eq:freeG})
\begin{equation}
 G_0(\bsy{x}, \bsy{x}') \stackrel{\bsy{x}\rightarrow\bsy{x}'}{\longrightarrow} -\frac{\imi}{2\pi} \frac{\bsy{\sigma}\cdot(\bsy{x}-\bsy{x}')}{|\bsy{x}-\bsy{x}'|^2}\,.
\end{equation}
If $\bsy{x}'$ lies in the interior of $\mathcal{V}$ and $\bsy{\alpha}$ is a point on the boundary $\partial \mathcal{V}$,
\begin{equation}
 \lim_{\bsy{x}\rightarrow \bsy{\alpha}} G_0(\bsy{x}, \bsy{x}') = G_0(\bsy{\alpha}, \bsy{x}')
\end{equation}
is well defined and the first term in Eq.\,(\ref{eq:jump}) is trivially obtained from Eq.\,(\ref{eq:FullG1}). However if $\bsy{x}'$ is on the boundary, the singular behavior of the Green's function becomes relevant. To see this, we perform the boundary integral in two parts, dividing $\partial \mathcal{V}$ into a small region
$D_{\delta}(\bsy{\alpha}) = C_{\delta}(\bsy{\alpha})\bigcap \partial \mathcal{V}$, where  $C_{\delta}(\bsy{\alpha})$ is a circle with radius $\delta$ around $\bsy{\alpha}$, and
the remaining border $\bar{D}_{\delta}(\bsy{\alpha}) = \partial \mathcal{V} \setminus D_{\delta}(\bsy{\alpha})$. We will take the limit $\delta\rightarrow 0$ at the end of the calculation.

We begin with the integration within $D_{\delta}(\bsy{\alpha})$. To this end we use the asymptotic expression for $G_0$ and get
\begin{eqnarray}
I_{D_{\delta}(\bsy{\alpha})} &=& \lim_{\delta\rightarrow 0}\lim_{\bsy{x}\rightarrow \bsy{\alpha}} \int\limits_{D_{\delta}(\bsy{\alpha})}\!\!\! d\sigma_{\bsy{\beta}} G_0(\bsy{x}, \bsy{\beta}) i \sigmav{\bsy{n_\beta}}\mu({\bsy{\beta}, \bsy{x}'})  \\
 &=& \frac{\sigmav{\bsy{n_\beta}}}{2\pi} \mu({\bsy{\alpha}, \bsy{x}'})\bsy{\sigma}\cdot \lim_{\delta\rightarrow 0}\lim_{\bsy{x}\rightarrow \bsy{\alpha}}
\int\limits_{D_{\delta}(\bsy{\alpha})} \!\!\!d\sigma_{\bsy{\beta}} \frac{(\bsy{x}-\bsy{\beta})}{|\bsy{x}-\bsy{\beta}|^2} ~, \nonumber
\end{eqnarray}
where we took $\mu$ out of the integral and evaluated it at $\bsy{\beta}=\bsy{\alpha}$. Without loss of generality, we choose $\bsy{\alpha}=\bsy{0}$, $\bsy{x}=|\bsy{x}|\hat{\bsy{y}}$ and approximate
$D_{\delta}(\bsy{\alpha})$ by a straight line along the $x$-axis, i.\,e. $D_{\delta}(\bsy{\alpha})~=~\{~\xi \hat{\bsy{x}}~|~ \xi \in [-\delta,\delta]~\}$.
Then we get
\begin{eqnarray}
  I_{D_{\delta}(\bsy{\alpha})} &=& \frac{\sigmav{\bsy{n_\beta}}}{2\pi} \mu({\bsy{\alpha}, \bsy{x}'})\bsy{\sigma}\cdot 
\lim_{\delta\rightarrow 0}\lim_{|\bsy{x}|\rightarrow 0}
\int\limits_{-\delta}^{\delta} d\xi \frac{|\bsy{x}|\hat{\bsy{y}}-\xi\hat{\bsy{x}}}{|\bsy{x}|^2+\xi^2} \nonumber \\
 &=&  
 \frac{\sigmav{\bsy{n_\beta}}}{2\pi} \mu({\bsy{\alpha}, \bsy{x}'})\bsy{\sigma}\cdot \lim_{\delta\rightarrow 0}\lim_{|\bsy{x}|\rightarrow 0}
  2 \arctan(\delta/|\bsy{x}|) \hat{\bsy{y}}\nonumber\\ 
&=& \frac{1}{2}\mu(\bsy{\alpha}, \bsy{x}') ~.
\end{eqnarray}
Since the kernel of the integral on $\bar{D}_{\delta}(\bsy{\alpha})$ has no singularity, it simply follows
\begin{eqnarray}
 \lim_{\delta\rightarrow 0}\lim\limits_{\bsy{x}\rightarrow \bsy{\alpha}}&& \!\!\!\!\!\!\!\!
 \int\limits_{\bar{D}_{\delta}(\bsy{\alpha})} d\sigma_{\bsy{\beta}} G_0(\bsy{x}, \bsy{\beta}) i\sigmav{\bsy{n_\beta}}\mu({\bsy{\beta}, \bsy{x}'}) \nonumber \\
 &=&\int\limits_{\partial V} d\sigma_{\bsy{\beta}} G_0(\bsy{\alpha}, \bsy{\beta}) i\sigmav{\bsy{n_\beta}}\mu({\bsy{\beta}, \bsy{x}'})~.
\end{eqnarray}
It is known from potential theory, that the integral on the right hand side exists\cite{Balian1970} and thus Eq.\,(\ref{eq:jump}) follows.


\section{Effective boundary condition for zigzag edges in the presence of next-nearest neighbor hopping}
\label{app:ZZNNN}

It has been shown in Refs. \onlinecite{Wimmer2008a} and \onlinecite{Sasaki2009} that the inclusion of next-nearest-neighbor (nnn) hopping in the tight-binding
Hamiltonian of graphene has important consequences on the properties of the zigzag edge states. While for bulk graphene, up to a constant energy shift, the effects are of subleading order in $k$, for finite samples nnn hopping leads to an additional effective potential that is located solely on the edge atoms, therefore leading to qualitative changes
of the edge state properties. These range from a finite dispersion to a complete change of the current profile in transport\cite{Wimmer2008a} . 

Here we neglect terms of higher order in $k$ in the Hamiltonian due to the nnn hopping and focus on the effects of the resulting edge potential. To this end we derive an effective boundary condition for the Dirac Hamiltonian with zigzag boundary. We consider a single zigzag edge, where the last row of atoms is located at $y_0 = a/\sqrt{3}$. Furthermore the graphene flake shall be extended for $y>y_0$, i.\,e. the last row of atoms is of $B$-type. The Hamiltonian is then given by\cite{Sasaki2009}
\begin{equation}
 \mathcal{H} = \vF \bsy{\sigma}\cdotsh \bsy{p} -\hbar \vF \frac{t'}{2} \delta(y-y_0) (1-\sigma_z\otimes \tau_z)\,.
\end{equation}
Here $t'\approx 0.1$ is the ratio of the next-nearest neighbor hopping constant, and the projection
$(1-\sigma_z\otimes \tau_z)$ ensures that the potential is located on the $B$-sublattice. Similar edge potentials can model also adsorbants at graphene edges
or edge magnetism \cite{Wimmer2010, Wimmer2008}.

The Dirac equation together with the Bloch theorem gives for the $y$-dependent part of the wavefunctions in the valley $\tau=+1$
\begin{eqnarray}
\label{eq:DEQsmally_a}
 \kE \psi_A(y) &=& k \psi_B(y) - \frac{\partial \psi_B(y)}{\partial y}\,, \\
\label{eq:DEQsmally_b}
 \kE \psi_B(y) &=& k \psi_B(y) + \frac{\partial \psi_A(y)}{\partial y} - t'\delta(y-y_0) \psi_B(y) \,. \nonumber \\
\end{eqnarray}
Now we integrate these equations over a small window $[y_0-\varepsilon, y_0+\varepsilon ]$ around the potential and take the limit $\varepsilon\rightarrow 0$ afterwards. Assuming that $\psi$ has at most a finite
discontinuity at $y_0$, we obtain from Eq.\,(\ref{eq:DEQsmally_a})
\begin{equation}
 \lim_{\varepsilon_0 \rightarrow 0^+} \psi_B(y+\varepsilon ) - \psi_B(y_0-\varepsilon ) = 0\,,
\end{equation}
i.\,e. the $B-$part of the spinor is continous. Thus we devide Eq.\,(\ref{eq:DEQsmally_b}) by $\psi_B(y)$ before integrating and get
\begin{eqnarray}
 t'&=& \lim_{\varepsilon \rightarrow 0^+} \intl_{y_0 - \varepsilon}^{y_0 + \varepsilon} \frac{1}{\psi_B(y)} \frac{\partial \psi_A(y)}{\partial y} \\
\label{eq:jumpNNN}
&=& \lim_{\varepsilon \rightarrow 0^+}\left[ \frac{\psi_A(y_0+\varepsilon)}{\psi_B(y_0+\varepsilon)} -\frac{\psi_A(y_0-\varepsilon)}{\psi_B(y_0-\varepsilon)}\right]
\end{eqnarray}
using integration by parts.
For $y<y_0$ we employ the actual zigzag boundary condition $\psi_A(0) = 0$,
leading to the known expressions for the wavefunctions for $y<y_0$\cite{Brey2006, Wurm2009a}:
\begin{eqnarray}
 \psi_{A}(y) &=& A \sin(qy)\,, \nonumber \\
 \psi_{B}(y) &=& \frac{A}{\kE} \left[ik\tau\sin(qy) + q\cos(qy)\right]\,,
\end{eqnarray}
with longitudinal and transverse momenta $k$ and $q$, respectively.
Since the effective Dirac equation is valid for momenta that are much smaller than $1/a$, we approximate $\kE a, q a, ka \approx 0$ to get
\begin{equation}
 \lim_{\varepsilon \rightarrow 0^+} \frac{\psi_A(y_0-\varepsilon)}{\psi_B(y_0-\varepsilon)} =
 \frac{\kE \sin(qa/\sqrt{3})}{ik \sin(qa/\sqrt{3}) +  q\cos(qa/\sqrt{3})} \approx 0\,,
\end{equation}
which inserted into Eq.\,(\ref{eq:jumpNNN}) finally leads to the effective boundary condition
\begin{equation}
  \left.\frac{\psi_A}{\psi_B}\right|_{\partial \mathcal{V}} = t'\,.
\end{equation}
in agreement with a result found for similar edge potentials in Ref. \onlinecite{Bhowmick2010}. In an analogous way one can derive the effective boundary condition for the other valley as well as for $A$-terminated zigzag edges to end up with an effective boundary condition matrix
\begin{equation}
 P_{\bsy{\alpha}} = \frac{1}{2} \left(1\mp \tau_z \otimes\sigma_z - it'\sigma_y \pm t'\tau_z\otimes\sigma_x\right)
\end{equation}
for all points at the edge $\bsy{\alpha}$. This expression turns into the usual zigzag matrix (\ref{eq:bczz_def}) when $t'=0$.

We further derive the edge state dispersion and wavefunction from the Dirac equation with the effective boundary condition
\begin{equation}
\label{eq:BC_wvfct}
 \psi_A(0) = t'\psi_B(0)\,.
\end{equation}
Due to the Bloch theorem we can write for $\kE^2 = k^2 +q^2$
\begin{equation}
 \Psi_A(x,y) = e^{ikx} \psi_A(y) = e^{ikx}(A e^{iqy} + B e^{-iqy})
\end{equation}
and
\begin{equation}
 \psi_B(y) = (\tau k+ \partial_y) \psi_A(y)\,.
\end{equation}
For nonzero $\kE$, Eq.\,(\ref{eq:BC_wvfct}) then leads to the condition
\begin{equation}
 A(\kE -t'\tau k-it' q) = -B(\kE -t'\tau k+it'q)\,.
\end{equation}
The bulk states result from this equation, when both sides are nonzero. On the other hand the edge state results if this is not the case, 
e.\,g. $\kE -t'\tau k+it' q = 0$. Solving this equation gives \textit{for negative $\tau k$} the edge state
\begin{equation}
 \Psi_A(x,y) \approx B e^{ikx}e^{\tau ky} \qquad \Psi_B(x,y) \approx B \frac{2k}{\kE} e^{ikx}e^{\tau ky}
\end{equation}
with the dispersion relation
\begin{equation}
\label{eq:dispersionNNN}
 \kE^{\text{edge}} = \frac{2k t'\tau}{1+t'^2} \approx 2k t'\tau\,.
\end{equation}
This state exists only for negative (positive) momenta $k$ in the valley $K(K')$ (as for the case without nnn hopping)
and has always a \textit{negative energy}.


\section{Energy eigenvalues of a rectangular graphene flake}
\label{app:Rectangle}

\label{app:Rectangle_ev}

Here we present an implicit expression for the energy eigenvalues of a graphene rectangle with zigzag edges at $y=0$ and $y=L_y$ and armchair edges at $x=0$ and \mbox{$x=L_x$}, respectively. To this end we start from a superposition of a forward and a backward propagating eigenmode of an armchair nanoribbon with edges at $x=0$ and $x=L_x$ \cite{Brey2006, Wurm2009a},
\begin{eqnarray}
\label{eq:Rect_wfct}
 \Psi(x,y)&=&
A\left(
\begin{array}{c}
 (q_m - ik)e^{iq_mx} \\
 \kE e^{iq_mx} \\
 \kE e^{-iq_mx} \\
 (-q_m + ik)e^{-iq_mx}
\end{array}
\right) e^{iky} \nonumber \\
&+&
B\left(
\begin{array}{c}
 (q_m + ik)e^{iq_mx} \\
 \kE e^{iq_mx} \\
 \kE e^{-iq_mx} \\
 (-q_m - ik)e^{-iq_mx}
\end{array}
\right) e^{-iky}
\end{eqnarray}
where the $q_m$ are quantized according to
\begin{equation}
q_m = \frac{m\pi}{L_x} - K \quad m \in \mathbb{Z}\,.
\end{equation}
The spinors in Eq.\,(\ref{eq:Rect_wfct}) are solutions to the Dirac equation when
\begin{equation}
\label{eq:Esq_ksq_qsq}
 k^2 + q_m^2 = \kE^2\,.
\end{equation}
Now we impose the zigzag boundary conditions $\Psi_A(x,0) = \Psi_{A'}(x,0) = \Psi_B(x,L_y) = \Psi_{B'}(x,L_y) =0$,
which result in the two independent equations
\begin{eqnarray}
 (q_m - ik)A +(q_m + ik)B &=& 0\,, \\
e^{ikL_y}A + e^{-ikL_y}B &=& 0\,.
\end{eqnarray}
These are solved for quantized $k_{nm}$ that fulfill the transcendental equation
\begin{equation}
\label{eq:quantiz_rect}
k_{nm} = - q_m \tan(k_{nm} L_y)\,.
\end{equation}
With that we have formally solved the problem, the eigenenergies
can be found e.\,g. by solving Eq.\,(\ref{eq:quantiz_rect}) numerically.

\section{Effect of weak bulk disorder }
\label{app:BulkDis}
At this point we briefly discuss the effect on the trace formula (\ref{eq:Gutzwiller}) caused by smooth bulk disorder, which can be accounted for by an additional term
\begin{equation}
 H' = \tau_0\otimes\sigma_0 V(\bsy{x})
\end{equation}
in the Hamiltonian, where $V(\bsy{x})$ is smooth on the scale of the lattice constant\cite{Suzuura2002}. In the semiclassical limit the Green's function for $H+H'$ has been derived in Ref. \onlinecite{Carmier2008} without taking into account the boundaries. 
For the case of a Gaussian correlated disorder potential,
\begin{equation}
 \left<V(\bsy{x})V(\bsy{x}')\right> = C_0 \exp\left[{-\frac{(\bsy{x}-\bsy{x}')^2}{4\Delta^2}}\right]\,,
\end{equation}
quantum calculations in the Boltzmann limit have been performed \cite{Adam2009, Vasko2010}. Under the assumption that the disorder potential is weak enough that the classical trajectories remain unaffected, we get for the impurity averaged Green's function\cite{Richter1996}
\begin{equation}
\label{eq:G0imp}
 \left<G_0'^{\text{sc}}(\bsy{x},\bsy{x}')\right> \approx G_0^{\text{sc}}(\bsy{x},\bsy{x}') \exp\left(-\left<\delta S^2\right>/2\hbar^2\right)
\end{equation}
with
\begin{eqnarray}
 \left<\delta S^2\right> &=& \frac{1}{(2\vF \kE)^2} \intl_{\bsy{x}'}^{\bsy{x}} dq \intl_{\bsy{x}'}^{\bsy{x}} dq'  
 \left< \left[ (\bsy{x}-\bsy{x}') \times \bsy{\nabla} V(\bsy{q}) \right]\right. \nonumber \\
 &&\times \left.\left[ (\bsy{x}-\bsy{x}') \times \bsy{\nabla} V(\bsy{q}') \right]\right>/ |\bsy{x}-\bsy{x}'|^2 \\ &\approx& \hbar^2|\bsy{x}-\bsy{x}'|/ l
\end{eqnarray}
and the mean free path
\begin{eqnarray}
 l = \frac{4 \Delta \hbar^2 \vF^2 \kE^2}{\sqrt{\pi} C_0}\,.
\end{eqnarray}
For smooth potentials the jump of the Green's function in Eq.\,(\ref{eq:jump}) and hence also the MRE (\ref{eq:fullG2_2}) remains unchanged, except that $G_0$ has to be replaced by its impurity averaged version (\ref{eq:G0imp}). Thus each summand in the semiclassical Green's function (\ref{eq:Gsc1}) for a graphene cavity aquires a damping factor 
$ e^{-L_\gamma/2l}$. In the trace integral (\ref{eq:traceint}), these factors do not alter the stationary phase points, so that also in the trace formula in (\ref{eq:Gutzwiller}) every periodic orbit contribution is weighted with a factor $ e^{-L_\gamma/2l}$ that improves convergence of the semiclassical trace formula.

%


\begin{thebibliography}{74}
\expandafter\ifx\csname natexlab\endcsname\relax\def\natexlab#1{#1}\fi
\expandafter\ifx\csname bibnamefont\endcsname\relax
  \def\bibnamefont#1{#1}\fi
\expandafter\ifx\csname bibfnamefont\endcsname\relax
  \def\bibfnamefont#1{#1}\fi
\expandafter\ifx\csname citenamefont\endcsname\relax
  \def\citenamefont#1{#1}\fi
\expandafter\ifx\csname url\endcsname\relax
  \def\url#1{\texttt{#1}}\fi
\expandafter\ifx\csname urlprefix\endcsname\relax\def\urlprefix{URL }\fi
\providecommand{\bibinfo}[2]{#2}
\providecommand{\eprint}[2][]{\url{#2}}

\bibitem[{\citenamefont{Novoselov et~al.}(2004)\citenamefont{Novoselov, Geim,
  Morozov, Jiang, Zhang, Dubonos, Grigorieva, and Firsov}}]{Novoselov2004}
\bibinfo{author}{\bibfnamefont{K.~S.} \bibnamefont{Novoselov}},
  \bibinfo{author}{\bibfnamefont{A.~K.} \bibnamefont{Geim}},
  \bibinfo{author}{\bibfnamefont{S.~V.} \bibnamefont{Morozov}},
  \bibinfo{author}{\bibfnamefont{D.}~\bibnamefont{Jiang}},
  \bibinfo{author}{\bibfnamefont{Y.}~\bibnamefont{Zhang}},
  \bibinfo{author}{\bibfnamefont{S.}~\bibnamefont{Dubonos}},
  \bibinfo{author}{\bibfnamefont{I.}~\bibnamefont{Grigorieva}},
  \bibnamefont{and} \bibinfo{author}{\bibfnamefont{A.}~\bibnamefont{Firsov}},
  \bibinfo{journal}{Science} \textbf{\bibinfo{volume}{306}},
  \bibinfo{pages}{666} (\bibinfo{year}{2004}).

\bibitem[{\citenamefont{Novoselov et~al.}(2005)\citenamefont{Novoselov, Jiang,
  Schedin, Booth, Khotkevich, Morozov, and Geim}}]{Novoselov2005}
\bibinfo{author}{\bibfnamefont{K.~S.} \bibnamefont{Novoselov}},
  \bibinfo{author}{\bibfnamefont{D.}~\bibnamefont{Jiang}},
  \bibinfo{author}{\bibfnamefont{F.}~\bibnamefont{Schedin}},
  \bibinfo{author}{\bibfnamefont{T.~J.} \bibnamefont{Booth}},
  \bibinfo{author}{\bibfnamefont{V.~V.} \bibnamefont{Khotkevich}},
  \bibinfo{author}{\bibfnamefont{S.~V.} \bibnamefont{Morozov}},
  \bibnamefont{and} \bibinfo{author}{\bibfnamefont{A.~K.} \bibnamefont{Geim}},
  \bibinfo{journal}{Proc. Natl. Acad. Sci. U.S.A.}
  \textbf{\bibinfo{volume}{102}}, \bibinfo{pages}{10451}
  (\bibinfo{year}{2005}).

\bibitem[{\citenamefont{Geim and Novoselov}(2007)}]{Geim2007}
\bibinfo{author}{\bibfnamefont{A.~K.} \bibnamefont{Geim}} \bibnamefont{and}
  \bibinfo{author}{\bibfnamefont{K.~S.} \bibnamefont{Novoselov}},
  \bibinfo{journal}{Nature Mater.} \textbf{\bibinfo{volume}{6}},
  \bibinfo{pages}{183 } (\bibinfo{year}{2007}).

\bibitem[{\citenamefont{Avouris et~al.}(2007)\citenamefont{Avouris, Chen, and
  Perebeinos}}]{Avouris2007}
\bibinfo{author}{\bibfnamefont{P.}~\bibnamefont{Avouris}},
  \bibinfo{author}{\bibfnamefont{Z.}~\bibnamefont{Chen}}, \bibnamefont{and}
  \bibinfo{author}{\bibfnamefont{V.}~\bibnamefont{Perebeinos}},
  \bibinfo{journal}{Nature Nanotech.} \textbf{\bibinfo{volume}{2}},
  \bibinfo{pages}{605 } (\bibinfo{year}{2007}).

\bibitem[{\citenamefont{Beenakker}(2008)}]{Beenakker2008}
\bibinfo{author}{\bibfnamefont{C.~W.~J.} \bibnamefont{Beenakker}},
  \bibinfo{journal}{Rev. Mod. Phys.} \textbf{\bibinfo{volume}{80}},
  \bibinfo{pages}{1337} (\bibinfo{year}{2008}).

\bibitem[{\citenamefont{Castro~Neto et~al.}(2009)\citenamefont{Castro~Neto,
  Guinea, Peres, Novoselov, and Geim}}]{Castro2009}
\bibinfo{author}{\bibfnamefont{A.~H.} \bibnamefont{Castro~Neto}},
  \bibinfo{author}{\bibfnamefont{F.}~\bibnamefont{Guinea}},
  \bibinfo{author}{\bibfnamefont{N.~M.~R.} \bibnamefont{Peres}},
  \bibinfo{author}{\bibfnamefont{K.~S.} \bibnamefont{Novoselov}},
  \bibnamefont{and} \bibinfo{author}{\bibfnamefont{A.~K.} \bibnamefont{Geim}},
  \bibinfo{journal}{Rev. Mod. Phys.} \textbf{\bibinfo{volume}{81}},
  \bibinfo{pages}{109} (\bibinfo{year}{2009}).

\bibitem[{\citenamefont{Abergel et~al.}(2010)\citenamefont{Abergel, Apalkov,
  Berashevich, Ziegler, and Chakraborty}}]{Abergel2010}
\bibinfo{author}{\bibfnamefont{D.}~\bibnamefont{Abergel}},
  \bibinfo{author}{\bibfnamefont{V.}~\bibnamefont{Apalkov}},
  \bibinfo{author}{\bibfnamefont{J.}~\bibnamefont{Berashevich}},
  \bibinfo{author}{\bibfnamefont{K.}~\bibnamefont{Ziegler}}, \bibnamefont{and}
  \bibinfo{author}{\bibfnamefont{T.}~\bibnamefont{Chakraborty}},
  \bibinfo{journal}{Adv. Phys.} \textbf{\bibinfo{volume}{59}},
  \bibinfo{pages}{261–482} (\bibinfo{year}{2010}).

\bibitem[{\citenamefont{Han et~al.}(2007)\citenamefont{Han, \"{O}zyilmaz,
  Zhang, and Kim}}]{Han2007}
\bibinfo{author}{\bibfnamefont{M.~Y.} \bibnamefont{Han}},
  \bibinfo{author}{\bibfnamefont{B.}~\bibnamefont{\"{O}zyilmaz}},
  \bibinfo{author}{\bibfnamefont{Y.}~\bibnamefont{Zhang}}, \bibnamefont{and}
  \bibinfo{author}{\bibfnamefont{P.}~\bibnamefont{Kim}},
  \bibinfo{journal}{Phys. Rev. Lett.} \textbf{\bibinfo{volume}{98}},
  \bibinfo{eid}{206805} (\bibinfo{year}{2007}).

\bibitem[{\citenamefont{Li et~al.}(2008)\citenamefont{Li, Wang, Zhang, Lee, and
  Dai}}]{Li2008}
\bibinfo{author}{\bibfnamefont{X.}~\bibnamefont{Li}},
  \bibinfo{author}{\bibfnamefont{X.}~\bibnamefont{Wang}},
  \bibinfo{author}{\bibfnamefont{L.}~\bibnamefont{Zhang}},
  \bibinfo{author}{\bibfnamefont{S.}~\bibnamefont{Lee}}, \bibnamefont{and}
  \bibinfo{author}{\bibfnamefont{H.}~\bibnamefont{Dai}},
  \bibinfo{journal}{Science} \textbf{\bibinfo{volume}{319}},
  \bibinfo{pages}{1229} (\bibinfo{year}{2008}).

\bibitem[{\citenamefont{Tapaszto et~al.}(2008)\citenamefont{Tapaszto, Dobrik,
  Lambin, and Biro}}]{Tapaszto2008}
\bibinfo{author}{\bibfnamefont{L.}~\bibnamefont{Tapaszto}},
  \bibinfo{author}{\bibfnamefont{G.}~\bibnamefont{Dobrik}},
  \bibinfo{author}{\bibfnamefont{P.}~\bibnamefont{Lambin}}, \bibnamefont{and}
  \bibinfo{author}{\bibfnamefont{L.}~\bibnamefont{Biro}},
  \bibinfo{journal}{Nature Nanotech.} \textbf{\bibinfo{volume}{3}},
  \bibinfo{pages}{397} (\bibinfo{year}{2008}).

\bibitem[{\citenamefont{Gallagher et~al.}(2010)\citenamefont{Gallagher, Todd,
  and Goldhaber-Gordon}}]{Gallagher2010}
\bibinfo{author}{\bibfnamefont{P.}~\bibnamefont{Gallagher}},
  \bibinfo{author}{\bibfnamefont{K.}~\bibnamefont{Todd}}, \bibnamefont{and}
  \bibinfo{author}{\bibfnamefont{D.}~\bibnamefont{Goldhaber-Gordon}},
  \bibinfo{journal}{Phys. Rev. B} \textbf{\bibinfo{volume}{81}},
  \bibinfo{pages}{115409} (\bibinfo{year}{2010}).

\bibitem[{\citenamefont{Ponomarenko et~al.}(2008)\citenamefont{Ponomarenko,
  Schedin, Katsnelson, Yang, Hill, Novoselov, and Geim}}]{Ponomarenko2008}
\bibinfo{author}{\bibfnamefont{L.~A.} \bibnamefont{Ponomarenko}},
  \bibinfo{author}{\bibfnamefont{F.}~\bibnamefont{Schedin}},
  \bibinfo{author}{\bibfnamefont{M.~I.} \bibnamefont{Katsnelson}},
  \bibinfo{author}{\bibfnamefont{R.}~\bibnamefont{Yang}},
  \bibinfo{author}{\bibfnamefont{E.~W.} \bibnamefont{Hill}},
  \bibinfo{author}{\bibfnamefont{K.~S.} \bibnamefont{Novoselov}},
  \bibnamefont{and} \bibinfo{author}{\bibfnamefont{A.~K.} \bibnamefont{Geim}},
  \bibinfo{journal}{Science} \textbf{\bibinfo{volume}{320}},
  \bibinfo{pages}{356} (\bibinfo{year}{2008}).

\bibitem[{\citenamefont{G\"uttinger et~al.}(2008)\citenamefont{G\"uttinger,
  Stampfer, Hellmüller, Molitor, Ihn, and Ensslin}}]{Guttinger2008}
\bibinfo{author}{\bibfnamefont{J.}~\bibnamefont{G\"uttinger}},
  \bibinfo{author}{\bibfnamefont{C.}~\bibnamefont{Stampfer}},
  \bibinfo{author}{\bibfnamefont{S.}~\bibnamefont{Hellm\"uller}},
  \bibinfo{author}{\bibfnamefont{F.}~\bibnamefont{Molitor}},
  \bibinfo{author}{\bibfnamefont{T.}~\bibnamefont{Ihn}}, \bibnamefont{and}
  \bibinfo{author}{\bibfnamefont{K.}~\bibnamefont{Ensslin}},
  \bibinfo{journal}{Appl. Phys. Lett.} \textbf{\bibinfo{volume}{93}},
  \bibinfo{pages}{212102} (\bibinfo{year}{2008}).


\bibitem[{\citenamefont{G\"uttinger et~al.}(2010)\citenamefont{G\"uttinger,
  Frey, Stampfer, Ihn, and Ensslin}}]{Guttinger2010}
\bibinfo{author}{\bibfnamefont{J.}~\bibnamefont{G\"uttinger}},
  \bibinfo{author}{\bibfnamefont{T.}~\bibnamefont{Frey}},
  \bibinfo{author}{\bibfnamefont{C.}~\bibnamefont{Stampfer}},
  \bibinfo{author}{\bibfnamefont{T.}~\bibnamefont{Ihn}}, \bibnamefont{and}
  \bibinfo{author}{\bibfnamefont{K.}~\bibnamefont{Ensslin}},
  \bibinfo{journal}{Phys. Rev. Lett.} \textbf{\bibinfo{volume}{105}},
  \bibinfo{pages}{116801} (\bibinfo{year}{2010}).


\bibitem{Russo2008}
  \bibinfo{author}{\bibfnamefont{S.}~\bibnamefont{Russo}},
  \bibinfo{author}{\bibfnamefont{J.~B.}~\bibnamefont{Oostinga}},
  \bibinfo{author}{\bibfnamefont{D.}~\bibnamefont{Wehenkel}},
  \bibinfo{author}{\bibfnamefont{H.~B.}~\bibnamefont{Heersche}},
  \bibinfo{author}{\bibfnamefont{S.~S.}~\bibnamefont{Sobhani}},
  \bibinfo{author}{\bibfnamefont{L.~M.~K.}~\bibnamefont{Vandersypen}}, \bibnamefont{and}
  \bibinfo{author}{\bibfnamefont{A.~F.}~\bibnamefont{Morpurgo}},
  \bibinfo{journal}{Phys. Rev. B} \textbf{\bibinfo{volume}{77}},
  \bibinfo{pages}{085413} (\bibinfo{year}{2008}).

\bibitem{Huefner2010}
  \bibinfo{author}{\bibfnamefont{M.}~\bibnamefont{Huefner}},
  \bibinfo{author}{\bibfnamefont{F.}~\bibnamefont{Molitor}},
  \bibinfo{author}{\bibfnamefont{A.}~\bibnamefont{Jacobsen}},
  \bibinfo{author}{\bibfnamefont{A.}~\bibnamefont{Pioda}},
  \bibinfo{author}{\bibfnamefont{C.}~\bibnamefont{Stampfer}},
  \bibinfo{author}{\bibfnamefont{K.}~\bibnamefont{Ensslin}}, \bibnamefont{and}
  \bibinfo{author}{\bibfnamefont{T.}~\bibnamefont{Ihn}},
  \bibinfo{journal}{New J. Phys} \textbf{\bibinfo{volume}{12}},
  \bibinfo{pages}{043054} (\bibinfo{year}{2010}).


\bibitem[{\citenamefont{Eroms and Weiss}(2009)}]{Eroms2009}
\bibinfo{author}{\bibfnamefont{J.}~\bibnamefont{Eroms}} \bibnamefont{and}
  \bibinfo{author}{\bibfnamefont{D.}~\bibnamefont{Weiss}},
  \bibinfo{journal}{New J. Phys.} \textbf{\bibinfo{volume}{11}},
  \bibinfo{pages}{095021} (\bibinfo{year}{2009}).

\bibitem[{\citenamefont{Bai et~al.}(2010)\citenamefont{Bai, Zhong, Jiang,
  Huang, and Duan}}]{Bai2010}
\bibinfo{author}{\bibfnamefont{J.}~\bibnamefont{Bai}},
  \bibinfo{author}{\bibfnamefont{X.}~\bibnamefont{Zhong}},
  \bibinfo{author}{\bibfnamefont{S.}~\bibnamefont{Jiang}},
  \bibinfo{author}{\bibfnamefont{Y.}~\bibnamefont{Huang}}, \bibnamefont{and}
  \bibinfo{author}{\bibfnamefont{X.}~\bibnamefont{Duan}},
  \bibinfo{journal}{Nature Nanotech.} \textbf{\bibinfo{volume}{5}},
  \bibinfo{pages}{190 } (\bibinfo{year}{2010}).


\bibitem[{\citenamefont{Fujita et~al.}(1996)\citenamefont{Fujita, Wakabayashi,
  Nakada, and Kusakabe}}]{Fujita1996}
\bibinfo{author}{\bibfnamefont{M.}~\bibnamefont{Fujita}},
  \bibinfo{author}{\bibfnamefont{K.}~\bibnamefont{Wakabayashi}},
  \bibinfo{author}{\bibfnamefont{K.}~\bibnamefont{Nakada}}, \bibnamefont{and}
  \bibinfo{author}{\bibfnamefont{K.}~\bibnamefont{Kusakabe}},
  \bibinfo{journal}{J. Phys. Soc. Jpn.} \textbf{\bibinfo{volume}{65}},
  \bibinfo{pages}{1920} (\bibinfo{year}{1996}).

\bibitem[{\citenamefont{Nakada et~al.}(1996)\citenamefont{Nakada, Fujita,
  Dresselhaus, and Dresselhaus}}]{Nakada1996}
\bibinfo{author}{\bibfnamefont{K.}~\bibnamefont{Nakada}},
  \bibinfo{author}{\bibfnamefont{M.}~\bibnamefont{Fujita}},
  \bibinfo{author}{\bibfnamefont{G.}~\bibnamefont{Dresselhaus}},
  \bibnamefont{and} \bibinfo{author}{\bibfnamefont{M.~S.}
  \bibnamefont{Dresselhaus}}, \bibinfo{journal}{Phys. Rev. B}
  \textbf{\bibinfo{volume}{54}}, \bibinfo{pages}{17954} (\bibinfo{year}{1996}).


\bibitem[{\citenamefont{Brey and Fertig}(2006{\natexlab{a}})}]{Brey2006}
\bibinfo{author}{\bibfnamefont{L.}~\bibnamefont{Brey}} \bibnamefont{and}
  \bibinfo{author}{\bibfnamefont{H.~A.} \bibnamefont{Fertig}},
  \bibinfo{journal}{Phys. Rev. B} \textbf{\bibinfo{volume}{73}},
  \bibinfo{pages}{235411} (\bibinfo{year}{2006}{\natexlab{a}}).

\bibitem{Tworzydlo2006}
  \bibinfo{author}{\bibfnamefont{J.}~\bibnamefont{Tworzyd\l o}},
  \bibinfo{author}{\bibfnamefont{B.}~\bibnamefont{Trauzettel}},
  \bibinfo{author}{\bibfnamefont{M.}~\bibnamefont{Titov}},
  \bibinfo{author}{\bibfnamefont{A.}~\bibnamefont{Rycerz}}, \bibnamefont{and}
  \bibinfo{author}{\bibfnamefont{C.~W.~J.}~\bibnamefont{Beenakker}},
  \bibinfo{journal}{Phys. Rev. Lett.} \textbf{\bibinfo{volume}{96}},
  \bibinfo{eid}{246802} (\bibinfo{year}{2006}).


\bibitem[{\citenamefont{Silvestrov and Efetov}(2007)}]{Silvestrov2007}
\bibinfo{author}{\bibfnamefont{P.~G.} \bibnamefont{Silvestrov}}
  \bibnamefont{and} \bibinfo{author}{\bibfnamefont{K.~B.}
  \bibnamefont{Efetov}}, \bibinfo{journal}{Phys. Rev. Lett.}
  \textbf{\bibinfo{volume}{98}}, \bibinfo{pages}{016802}
  (\bibinfo{year}{2007}).

\bibitem[{\citenamefont{Trauzettel et~al.}(2007)\citenamefont{Trauzettel,
  Bulaev, Loss, and Burkard}}]{Trauzettel2007}
\bibinfo{author}{\bibfnamefont{B.}~\bibnamefont{Trauzettel}},
  \bibinfo{author}{\bibfnamefont{D.~V.} \bibnamefont{Bulaev}},
  \bibinfo{author}{\bibfnamefont{D.}~\bibnamefont{Loss}}, \bibnamefont{and}
  \bibinfo{author}{\bibfnamefont{G.}~\bibnamefont{Burkard}},
  \bibinfo{journal}{Nature Phys.} \textbf{\bibinfo{volume}{3}},
  \bibinfo{pages}{192} (\bibinfo{year}{2007}).

\bibitem{Recher2009}
\bibinfo{author}{\bibfnamefont{P.}~\bibnamefont{Recher}},
\bibinfo{author}{\bibfnamefont{J.}~\bibnamefont{Nilsson}},
\bibinfo{author}{\bibfnamefont{G.}~\bibnamefont{Burkard}}, \bibnamefont{and}
\bibinfo{author}{\bibfnamefont{B.}~\bibnamefont{Trauzettel}},
  \bibinfo{journal}{Phys. Rev. B} \textbf{\bibinfo{volume}{79}},
  \bibinfo{pages}{085407} (\bibinfo{year}{2009}).

\bibitem{Bardarson2009}
\bibinfo{author}{\bibfnamefont{J.~H.}~\bibnamefont{Bardarson}},
\bibinfo{author}{\bibfnamefont{M.}~\bibnamefont{Titov}}, \bibnamefont{and}
\bibinfo{author}{\bibfnamefont{P.~W.}~\bibnamefont{Brouwer}},
  \bibinfo{journal}{Phys. Rev. Lett.} \textbf{\bibinfo{volume}{102}},
  \bibinfo{pages}{226803} (\bibinfo{year}{2009}).

\bibitem[{\citenamefont{Libisch et~al.}(2009)\citenamefont{Libisch, Stampfer,
  and Burgd\"orfer}}]{Libisch2009}
\bibinfo{author}{\bibfnamefont{F.}~\bibnamefont{Libisch}},
  \bibinfo{author}{\bibfnamefont{C.}~\bibnamefont{Stampfer}}, \bibnamefont{and}
  \bibinfo{author}{\bibfnamefont{J.}~\bibnamefont{Burgd\"orfer}},
  \bibinfo{journal}{Phys. Rev. B} \textbf{\bibinfo{volume}{79}},
  \bibinfo{pages}{115423} (\bibinfo{year}{2009}).

\bibitem[{\citenamefont{Wurm et~al.}(2009{\natexlab{a}})\citenamefont{Wurm,
  Rycerz, Adagideli, Wimmer, Richter, and Baranger}}]{Wurm2009}
\bibinfo{author}{\bibfnamefont{J.}~\bibnamefont{Wurm}},
  \bibinfo{author}{\bibfnamefont{A.}~\bibnamefont{Rycerz}},
  \bibinfo{author}{\bibfnamefont{\.{I}.} \bibnamefont{Adagideli}},
  \bibinfo{author}{\bibfnamefont{M.}~\bibnamefont{Wimmer}},
  \bibinfo{author}{\bibfnamefont{K.}~\bibnamefont{Richter}}, \bibnamefont{and}
  \bibinfo{author}{\bibfnamefont{H.~U.} \bibnamefont{Baranger}},
  \bibinfo{journal}{Phys. Rev. Lett.} \textbf{\bibinfo{volume}{102}},
  \bibinfo{pages}{056806} (\bibinfo{year}{2009}{\natexlab{a}}).


\bibitem[{\citenamefont{Wimmer et~al.}(2010)\citenamefont{Wimmer, Akhmerov, and
  Guinea}}]{Wimmer2010}
\bibinfo{author}{\bibfnamefont{M.}~\bibnamefont{Wimmer}},
  \bibinfo{author}{\bibfnamefont{A.~R.} \bibnamefont{Akhmerov}},
  \bibnamefont{and} \bibinfo{author}{\bibfnamefont{F.}~\bibnamefont{Guinea}},
  \bibinfo{journal}{Phys. Rev. B} \textbf{\bibinfo{volume}{82}},
  \bibinfo{pages}{045409} (\bibinfo{year}{2010}).

\bibitem{Recher2007}
  \bibinfo{author}{\bibfnamefont{P.}~\bibnamefont{Recher}},
  \bibinfo{author}{\bibfnamefont{B.}~\bibnamefont{Trauzettel}},
  \bibinfo{author}{\bibfnamefont{A.}~\bibnamefont{Rycerz}},
  \bibinfo{author}{\bibfnamefont{Ya.~M.}~\bibnamefont{Blanter}},
  \bibinfo{author}{\bibfnamefont{C.~W.~J.}~\bibnamefont{Beenakker}}, \bibnamefont{and} 
  \bibinfo{author}{\bibfnamefont{A.~F.}~\bibnamefont{Morpurgo}},
  \bibinfo{journal}{Phys. Rev. B} \textbf{\bibinfo{volume}{76}},
  \bibinfo{pages}{235404} (\bibinfo{year}{2007}).

\bibitem{Wurm2010}
  \bibinfo{author}{\bibfnamefont{J.}~\bibnamefont{Wurm}},
  \bibinfo{author}{\bibfnamefont{M.}~\bibnamefont{Wimmer}},
  \bibinfo{author}{\bibfnamefont{H.~U.}~\bibnamefont{Baranger}}, \bibnamefont{and}
  \bibinfo{author}{\bibfnamefont{K.}~\bibnamefont{Richter}},
  \bibinfo{journal}{Semicond. Sci. Technol.} \textbf{\bibinfo{volume}{25}},
  \bibinfo{pages}{034003} (\bibinfo{year}{2010}).

\bibitem[{\citenamefont{Schelter et~al.}(2010)\citenamefont{Schelter, Bohr, and
  Trauzettel}}]{Schelter2010}
\bibinfo{author}{\bibfnamefont{J.}~\bibnamefont{Schelter}},
  \bibinfo{author}{\bibfnamefont{D.}~\bibnamefont{Bohr}}, \bibnamefont{and}
  \bibinfo{author}{\bibfnamefont{B.}~\bibnamefont{Trauzettel}},
  \bibinfo{journal}{Phys. Rev. B} \textbf{\bibinfo{volume}{81}},
  \bibinfo{pages}{195441} (\bibinfo{year}{2010}).


\bibitem[{\citenamefont{Pedersen et~al.}(2008)\citenamefont{Pedersen, Flindt,
  Pedersen, Mortensen, Jauho, and Pedersen}}]{Pedersen2008}
\bibinfo{author}{\bibfnamefont{T.~G.} \bibnamefont{Pedersen}},
  \bibinfo{author}{\bibfnamefont{C.}~\bibnamefont{Flindt}},
  \bibinfo{author}{\bibfnamefont{J.}~\bibnamefont{Pedersen}},
  \bibinfo{author}{\bibfnamefont{N.~A.} \bibnamefont{Mortensen}},
  \bibinfo{author}{\bibfnamefont{A.-P.} \bibnamefont{Jauho}}, \bibnamefont{and}
  \bibinfo{author}{\bibfnamefont{K.}~\bibnamefont{Pedersen}},
  \bibinfo{journal}{Phys. Rev. Lett.} \textbf{\bibinfo{volume}{100}},
  \bibinfo{pages}{136804} (\bibinfo{year}{2008}).

\bibitem[{\citenamefont{Vanevi\ifmmode~\acute{c}\else \'{c}\fi{}
  et~al.}(2009)\citenamefont{Vanevi\ifmmode~\acute{c}\else \'{c}\fi{},
  Stojanovi\ifmmode~\acute{c}\else \'{c}\fi{}, and Kindermann}}]{Vanevic2009}
\bibinfo{author}{\bibfnamefont{M.}~\bibnamefont{Vanevi\ifmmode~\acute{c}\else
  \'{c}\fi{}}}, \bibinfo{author}{\bibfnamefont{V.~M.}
  \bibnamefont{Stojanovi\ifmmode~\acute{c}\else \'{c}\fi{}}}, \bibnamefont{and}
  \bibinfo{author}{\bibfnamefont{M.}~\bibnamefont{Kindermann}},
  \bibinfo{journal}{Phys. Rev. B} \textbf{\bibinfo{volume}{80}},
  \bibinfo{pages}{045410} (\bibinfo{year}{2009}).
  
\bibitem{Herrera2010}
\bibinfo{author}{\bibfnamefont{W.~J.}~\bibnamefont{Herrera}},
  \bibinfo{author}{\bibfnamefont{P.} \bibnamefont{Burset}}, \bibnamefont{and}
    \bibinfo{author}{\bibfnamefont{A.~L.} \bibnamefont{Yeyati}}, 
  \bibinfo{journal}{J. Phys.: Condens. Matter} \textbf{\bibinfo{volume}{22}},
  \bibinfo{pages}{275304} (\bibinfo{year}{2010}).


\bibitem[{\citenamefont{McCann and Fal'ko}(2004)}]{McCann2004}
\bibinfo{author}{\bibfnamefont{E.}~\bibnamefont{McCann}} \bibnamefont{and}
  \bibinfo{author}{\bibfnamefont{V.~I.} \bibnamefont{Fal'ko}},
  \bibinfo{journal}{J. Phys.: Condens. Matter} \textbf{\bibinfo{volume}{16}},
  \bibinfo{pages}{2371} (\bibinfo{year}{2004}).

\bibitem[{\citenamefont{Akhmerov and Beenakker}(2007)}]{Akhmerov2007}
\bibinfo{author}{\bibfnamefont{A.~R.} \bibnamefont{Akhmerov}} \bibnamefont{and}
  \bibinfo{author}{\bibfnamefont{C.~W.~J.} \bibnamefont{Beenakker}},
  \bibinfo{journal}{Phys. Rev. Lett.} \textbf{\bibinfo{volume}{98}},
  \bibinfo{pages}{157003} (\bibinfo{year}{2007}).

\bibitem[{\citenamefont{Akhmerov and Beenakker}(2008)}]{Akhmerov2008}
\bibinfo{author}{\bibfnamefont{A.~R.} \bibnamefont{Akhmerov}} \bibnamefont{and}
  \bibinfo{author}{\bibfnamefont{C.~W.~J.} \bibnamefont{Beenakker}},
  \bibinfo{journal}{Phys. Rev. B} \textbf{\bibinfo{volume}{77}},
  \bibinfo{eid}{085423} (\bibinfo{year}{2008}).

\bibitem[{\citenamefont{Balian and Bloch}(1970)}]{Balian1970}
\bibinfo{author}{\bibfnamefont{R.}~\bibnamefont{Balian}} \bibnamefont{and}
  \bibinfo{author}{\bibfnamefont{C.}~\bibnamefont{Bloch}},
  \bibinfo{journal}{Ann. Phys.} \textbf{\bibinfo{volume}{60}},
  \bibinfo{pages}{401} (\bibinfo{year}{1970}).
  
  \bibitem[{\citenamefont{Adagideli and Goldbart}(2002)}]{Adagideli2002}
\bibinfo{author}{\bibfnamefont{I.}~\bibnamefont{Adagideli}} \bibnamefont{and}
  \bibinfo{author}{\bibfnamefont{P.~M.} \bibnamefont{Goldbart}},
  \bibinfo{journal}{Int. J. Mod. Phys. B} \textbf{\bibinfo{volume}{16}},
  \bibinfo{pages}{1381} (\bibinfo{year}{2002}).
  
  \bibitem[{\citenamefont{Weyl}(1911)}]{Weyl1911}
\bibinfo{author}{\bibfnamefont{H.}~\bibnamefont{Weyl}},
  \bibinfo{journal}{Nachr. Akad. Wiss. Goettingen} p. \bibinfo{pages}{110}
  (\bibinfo{year}{1911}).

\bibitem[{\citenamefont{Gutzwiller}(1990)}]{Gutzwiller1990}
\bibinfo{author}{\bibfnamefont{M.~C.}~\bibnamefont{Gutzwiller}},
  \emph{\bibinfo{title}{Chaos in Classical and Quantum Mechanics}}
  (\bibinfo{publisher}{Springer}, \bibinfo{address}{New York},
  \bibinfo{year}{1990}).
  
\bibitem{Cserti2004}
\bibinfo{author}{\bibfnamefont{J.}~\bibnamefont{Cserti}},
\bibinfo{author}{\bibfnamefont{A.}~\bibnamefont{Csord\'{a}s}}, \bibnamefont{and}
  \bibinfo{author}{\bibfnamefont{U.}~\bibnamefont{Z\"ulicke}},
  \bibinfo{journal}{Phys. Rev. B} \textbf{\bibinfo{volume}{70}},
  \bibinfo{pages}{233307} (\bibinfo{year}{2004}).
  
\bibitem{Berry1976/77}
\bibinfo{author}{\bibfnamefont{M.~V.}~\bibnamefont{Berry}} \bibnamefont{and}
  \bibinfo{author}{\bibfnamefont{M.}~\bibnamefont{Tabor}},
  \bibinfo{journal}{Proc. Roy. Soc. Lond.} \textbf{\bibinfo{volume}{349}},
  \bibinfo{pages}{101} (\bibinfo{year}{1976}).\\
\bibinfo{author}{\bibfnamefont{M.~V.}~\bibnamefont{Berry}} \bibnamefont{and}
  \bibinfo{author}{\bibfnamefont{M.}~\bibnamefont{Tabor}},
  \bibinfo{journal}{J. Phys. A} \textbf{\bibinfo{volume}{10}},
  \bibinfo{pages}{371} (\bibinfo{year}{1977}).

\bibitem{Bolte1999}
\bibinfo{author}{\bibfnamefont{J.}~\bibnamefont{Bolte}} \bibnamefont{and}
  \bibinfo{author}{\bibfnamefont{S.} \bibnamefont{Keppeler}},
  \bibinfo{journal}{Ann. Phys.} \textbf{\bibinfo{volume}{274}},
  \bibinfo{pages}{125} (\bibinfo{year}{1999}).

\bibitem[{\citenamefont{Carmier and Ullmo}(2008)}]{Carmier2008}
\bibinfo{author}{\bibfnamefont{P.}~\bibnamefont{Carmier}} \bibnamefont{and}
  \bibinfo{author}{\bibfnamefont{D.}~\bibnamefont{Ullmo}},
  \bibinfo{journal}{Phys. Rev. B} \textbf{\bibinfo{volume}{77}},
  \bibinfo{pages}{245413} (\bibinfo{year}{2008}).

\bibitem{Pletyukhov2002}
  \bibinfo{author}{\bibfnamefont{M.}~\bibnamefont{Pletyukhov}},
  \bibinfo{author}{\bibfnamefont{Ch.}~\bibnamefont{Amann}},
\bibinfo{author}{\bibfnamefont{M.}~\bibnamefont{Mehta}}, \bibnamefont{and}
  \bibinfo{author}{\bibfnamefont{M.}~\bibnamefont{Brack}},
  \bibinfo{journal}{Phys. Rev. Lett.} \textbf{\bibinfo{volume}{89}},
  \bibinfo{pages}{116601} (\bibinfo{year}{2002}).
  
  \bibitem{Chang2004}
  \bibinfo{author}{\bibfnamefont{C.-H.}~\bibnamefont{Chang}},
  \bibinfo{author}{\bibfnamefont{A.~G.}~\bibnamefont{Mal'shukov}}, \bibnamefont{and}
  \bibinfo{author}{\bibfnamefont{K.~A.}~\bibnamefont{Chao}},
  \bibinfo{journal}{Phys. Rev. B} \textbf{\bibinfo{volume}{70}},
  \bibinfo{pages}{245309} (\bibinfo{year}{2004}).
  
\bibitem{Zaitsev2005}
  \bibinfo{author}{\bibfnamefont{O.}~\bibnamefont{Zaitsev}},
  \bibinfo{author}{\bibfnamefont{D.}~\bibnamefont{Frustaglia}},\bibnamefont{and}
  \bibinfo{author}{\bibfnamefont{K.}~\bibnamefont{Richter}},
  \bibinfo{journal}{Phys. Rev. B} \textbf{\bibinfo{volume}{72}},
  \bibinfo{pages}{155325} (\bibinfo{year}{2005})

\bibitem{Adagideli2010}
  \bibinfo{author}{\bibfnamefont{\.{I}.}~\bibnamefont{Adagideli}},
  \bibinfo{author}{\bibfnamefont{Ph.}~\bibnamefont{Jacquod}},
  \bibinfo{author}{\bibfnamefont{M.}~\bibnamefont{Scheid}},
  \bibinfo{author}{\bibfnamefont{M.}~\bibnamefont{Duckheim}},
  \bibinfo{author}{\bibfnamefont{D.}~\bibnamefont{Loss}}, \bibnamefont{and}
  \bibinfo{author}{\bibfnamefont{K.}~\bibnamefont{Richter}},
  \bibinfo{journal}{Phys. Rev. Lett.} \textbf{\bibinfo{volume}{105}},
  \bibinfo{pages}{246807} (\bibinfo{year}{2010})
  
  \bibitem{Kormanyos2008}
  \bibinfo{author}{\bibfnamefont{A.}~\bibnamefont{Korm\'{a}nyos}},
  \bibinfo{author}{\bibfnamefont{P.}~\bibnamefont{Rakyta}},
  \bibinfo{author}{\bibfnamefont{L.}~\bibnamefont{Oroszl\'{a}ny}}, \bibnamefont{and}
  \bibinfo{author}{\bibfnamefont{J.}~\bibnamefont{Cserti}},
  \bibinfo{journal}{Phys. Rev. B} \textbf{\bibinfo{volume}{78}},
  \bibinfo{pages}{045430} (\bibinfo{year}{2008})
  
    \bibitem{Rakyta2010}  
    \bibinfo{author}{\bibfnamefont{P.}~\bibnamefont{Rakyta}},
  \bibinfo{author}{\bibfnamefont{A.}~\bibnamefont{Korm\'{a}nyos}},
  \bibinfo{author}{\bibfnamefont{J.}~\bibnamefont{Cserti}}, \bibnamefont{and}
\bibinfo{author}{\bibfnamefont{P}~\bibnamefont{Koskinen}},
  \bibinfo{journal}{Phys. Rev. B} \textbf{\bibinfo{volume}{81}},
  \bibinfo{pages}{115411} (\bibinfo{year}{201ß})
  
  \bibitem{Carmier2010}
  \bibinfo{author}{\bibfnamefont{P.}~\bibnamefont{Carmier}},
\bibinfo{author}{\bibfnamefont{C.}~\bibnamefont{Lewenkopf}}, \bibnamefont{and}
  \bibinfo{author}{\bibfnamefont{D.}~\bibnamefont{Ullmo}},
  \bibinfo{journal}{Phys. Rev. B} \textbf{\bibinfo{volume}{81}},
  \bibinfo{pages}{241406} (\bibinfo{year}{2010}).

\bibitem[{\citenamefont{Wallace}(1947)}]{Wallace1947}
\bibinfo{author}{\bibfnamefont{P.~R.} \bibnamefont{Wallace}},
  \bibinfo{journal}{Phys. Rev.} \textbf{\bibinfo{volume}{71}},
  \bibinfo{pages}{622} (\bibinfo{year}{1947}).

\bibitem{note_1}
{This definition is, for graphene, related to the common definition $DOS(E) = \sum_n \delta(E-E_n)$ via \mbox{$\rho(\kE)=\hbar \vF DOS(E)$}.}

\bibitem[{\citenamefont{Brack and Bhaduri}(2008)}]{Brack2008}
\bibinfo{author}{\bibfnamefont{M.}~\bibnamefont{Brack}} \bibnamefont{and}
  \bibinfo{author}{\bibfnamefont{R.}~\bibnamefont{Bhaduri}},
  \emph{\bibinfo{title}{Semiclassical Physics}}
  (\bibinfo{publisher}{Addison-Wesley, New York}, \bibinfo{year}{2008}).

\bibitem[{\citenamefont{St\"ockmann}(1999)}]{Stockmann1999}
\bibinfo{author}{\bibfnamefont{H.-J.} \bibnamefont{St\"ockmann}},
  \emph{\bibinfo{title}{Quantum Chaos}} (\bibinfo{publisher}{Cambridge Univ.
  Press}, \bibinfo{address}{Cambridge}, \bibinfo{year}{1999}).

\bibitem[{\citenamefont{Wimmer}(2008)}]{Wimmer2008a}
\bibinfo{author}{\bibfnamefont{M.}~\bibnamefont{Wimmer}}, Ph.D. thesis,
  \bibinfo{school}{Universit\"at Regensburg} (\bibinfo{year}{2008}).

\bibitem[{\citenamefont{Kobayashi et~al.}(2005)\citenamefont{Kobayashi, Fukui,
  Enoki, Kusakabe, and Kaburagi}}]{Kobayashi2005}
\bibinfo{author}{\bibfnamefont{Y.}~\bibnamefont{Kobayashi}},
  \bibinfo{author}{\bibfnamefont{K.-I
.} \bibnamefont{Fukui}},
  \bibinfo{author}{\bibfnamefont{T.}~\bibnamefont{Enoki}},
  \bibinfo{author}{\bibfnamefont{K.}~\bibnamefont{Kusakabe}}, \bibnamefont{and}
  \bibinfo{author}{\bibfnamefont{Y.}~\bibnamefont{Kaburagi}},
  \bibinfo{journal}{Phys. Rev. B} \textbf{\bibinfo{volume}{71}},
  \bibinfo{pages}{193406} (\bibinfo{year}{2005}).

\bibitem[{\citenamefont{Niimi et~al.}(2006)\citenamefont{Niimi, Matsui,
  Kambara, Tagami, Tsukada, and Fukuyama}}]{Niimi2006}
\bibinfo{author}{\bibfnamefont{Y.}~\bibnamefont{Niimi}},
  \bibinfo{author}{\bibfnamefont{T.}~\bibnamefont{Matsui}},
  \bibinfo{author}{\bibfnamefont{H.}~\bibnamefont{Kambara}},
  \bibinfo{author}{\bibfnamefont{K.}~\bibnamefont{Tagami}},
  \bibinfo{author}{\bibfnamefont{M.}~\bibnamefont{Tsukada}}, \bibnamefont{and}
  \bibinfo{author}{\bibfnamefont{H.}~\bibnamefont{Fukuyama}},
  \bibinfo{journal}{Phys. Rev. B} \textbf{\bibinfo{volume}{73}},
  \bibinfo{pages}{085421} (\bibinfo{year}{2006}).

\bibitem[{\citenamefont{Sasaki et~al.}(2009)\citenamefont{Sasaki, Shimomura,
  Takane, and Wakabayashi}}]{Sasaki2009}
\bibinfo{author}{\bibfnamefont{K.~I.}~\bibnamefont{Sasaki}},
  \bibinfo{author}{\bibfnamefont{Y.}~\bibnamefont{Shimomura}},
  \bibinfo{author}{\bibfnamefont{Y.}~\bibnamefont{Takane}}, \bibnamefont{and}
  \bibinfo{author}{\bibfnamefont{K.}~\bibnamefont{Wakabayashi}},
  \bibinfo{journal}{Phys. Rev. Lett.} \textbf{\bibinfo{volume}{102}},
  \bibinfo{pages}{146806} (\bibinfo{year}{2009}).

\bibitem{note_2}
{$\sigma_z$ commutes with the effective Hamiltonian without nnn hopping, giving rise to a particle hole symmetry, which is broken by the nnn-term at the edges (cf. appendix \ref{app:ZZNNN}).}

\bibitem[{\citenamefont{Berry and Mondragon}(1987)}]{Berry1987}
\bibinfo{author}{\bibfnamefont{M.}~\bibnamefont{Berry}} \bibnamefont{and}
  \bibinfo{author}{\bibfnamefont{R.}~\bibnamefont{Mondragon}},
  \bibinfo{journal}{Proc. R. Soc. Lond. A} \textbf{\bibinfo{volume}{412}},
  \bibinfo{pages}{53} (\bibinfo{year}{1987}).

\bibitem[{\citenamefont{Wimmer and Richter}(2009)}]{Wimmer2009}
\bibinfo{author}{\bibfnamefont{M.}~\bibnamefont{Wimmer}} \bibnamefont{and}
  \bibinfo{author}{\bibfnamefont{K.}~\bibnamefont{Richter}},
  \bibinfo{journal}{J. Comp. Phys.} \textbf{\bibinfo{volume}{228}},
  \bibinfo{pages}{8548 } (\bibinfo{year}{2009}).

\bibitem[{\citenamefont{Balian and Bloch}(1972)}]{Balian1972}
\bibinfo{author}{\bibfnamefont{R.}~\bibnamefont{Balian}} \bibnamefont{and}
  \bibinfo{author}{\bibfnamefont{C.}~\bibnamefont{Bloch}},
  \bibinfo{journal}{Ann. Phys.} \textbf{\bibinfo{volume}{69}},
  \bibinfo{pages}{76} (\bibinfo{year}{1972}).

\bibitem{Bleistein1975}
\bibinfo{author}{\bibfnamefont{N.}~\bibnamefont{Bleistein}} \bibnamefont{and}
  \bibinfo{author}{\bibfnamefont{R.~A.}~\bibnamefont{Handelsmann}},
  \emph{\bibinfo{title}{Asymptotic expansions of integrals}}
  (\bibinfo{publisher}{Rinehart and Winston, New York}, \bibinfo{year}{1975}).

\bibitem[{\citenamefont{Baranger et~al.}(1993)\citenamefont{Baranger, Jalabert,
  and Stone}}]{Baranger1993a}
\bibinfo{author}{\bibfnamefont{H.~U.} \bibnamefont{Baranger}},
  \bibinfo{author}{\bibfnamefont{R.~A.} \bibnamefont{Jalabert}},
  \bibnamefont{and} \bibinfo{author}{\bibfnamefont{A.~D.} \bibnamefont{Stone}},
  \bibinfo{journal}{Chaos} \textbf{\bibinfo{volume}{3}}, \bibinfo{pages}{665}
  (\bibinfo{year}{1993}).

\bibitem[{\citenamefont{Berry}(1985)}]{Berry1985}
\bibinfo{author}{\bibfnamefont{M.}~\bibnamefont{Berry}},
  \bibinfo{journal}{Proc. R. Soc. Lond. A} \textbf{\bibinfo{volume}{400}},
  \bibinfo{pages}{229} (\bibinfo{year}{1985}).

\bibitem[{\citenamefont{Bogachek and Gogadze}(1973)}]{Bogachek1973}
\bibinfo{author}{\bibfnamefont{E.}~\bibnamefont{Bogachek}}, \bibnamefont{and}
  \bibinfo{author}{\bibfnamefont{G.}~\bibnamefont{Gogadze}},
  \bibinfo{journal}{Sov. Phys. JETP} \textbf{\bibinfo{volume}{36}},
  \bibinfo{pages}{973} (\bibinfo{year}{1973}).
%

\bibitem{Richter1996}
  \bibinfo{author}{\bibfnamefont{K.}~\bibnamefont{Richter}},
  \bibinfo{author}{\bibfnamefont{D.}~\bibnamefont{Ullmo}}, \bibnamefont{and}
  \bibinfo{author}{\bibfnamefont{R.~A.}~\bibnamefont{Jalabert}},
  \bibinfo{journal}{Phys. Rev. B} \textbf{\bibinfo{volume}{54}},
  \bibinfo{pages}{5219} (\bibinfo{year}{1996}).



\bibitem[{\citenamefont{Richter et~al.}(1996)\citenamefont{Richter, Ullmo, and
  Jalabert}}]{Richter1996a}
\bibinfo{author}{\bibfnamefont{K.}~\bibnamefont{Richter}},
  \bibinfo{author}{\bibfnamefont{D.}~\bibnamefont{Ullmo}}, \bibnamefont{and}
  \bibinfo{author}{\bibfnamefont{R.}~\bibnamefont{Jalabert}},
  \bibinfo{journal}{Phys. Rep.} \textbf{\bibinfo{volume}{276}},
  \bibinfo{pages}{1} (\bibinfo{year}{1996}).

\bibitem{note_3}
{Armchair graphene nanoribbons with such a width are metallic, while all others are semiconducting.}

\bibitem{Suzuura2002}
\bibinfo{author}{\bibfnamefont{H.}~\bibnamefont{Suzuura}}, \bibnamefont{and}
\bibinfo{author}{\bibfnamefont{T.}~\bibnamefont{Ando}},
\bibinfo{journal}{Phys. Rev. Lett.} \textbf{\bibinfo{volume}{89}},
\bibinfo{pages}{266603} (\bibinfo{year}{2002}).

\bibitem{Adam2009}
\bibinfo{author}{\bibfnamefont{S.}~\bibnamefont{Adam}},
\bibinfo{author}{\bibfnamefont{P.~W.}~\bibnamefont{Brouwer}}, \bibnamefont{and}
\bibinfo{author}{\bibfnamefont{S.}~\bibnamefont{Das Sarma}},
\bibinfo{journal}{Phys. Rev. B} \textbf{\bibinfo{volume}{79}},
\bibinfo{pages}{201404} (\bibinfo{year}{2009}).

\bibitem{Vasko2010}
\bibinfo{author}{\bibfnamefont{F.~T.}~\bibnamefont{Vasko}}, \bibnamefont{and}
\bibinfo{author}{\bibfnamefont{I.~V.}~\bibnamefont{Zozoulenko}}, 
\bibinfo{journal}{App. Phys. Lett.} \textbf{\bibinfo{volume}{97}},
\bibinfo{pages}{092115} (\bibinfo{year}{2010}).
  
\bibitem{partII}
\bibinfo{author}{\bibfnamefont{J.}~\bibnamefont{Wurm}},
\bibinfo{author}{\bibfnamefont{\.{I}.}~\bibnamefont{Adagideli}}, and
\bibinfo{author}{\bibfnamefont{K.}~\bibnamefont{Richter}},
\bibinfo{journal}{to be submitted to Phys. Rev. B.}

\bibitem{Son2006}
\bibinfo{author}{\bibfnamefont{Y.-W.}~\bibnamefont{Son}},
\bibinfo{author}{\bibfnamefont{M.~L.}~\bibnamefont{Cohen}}, \bibnamefont{and}
\bibinfo{author}{\bibfnamefont{S.~G.}~\bibnamefont{Louie}},
  \bibinfo{journal}{Nature} \textbf{\bibinfo{volume}{444}},
  \bibinfo{pages}{347} (\bibinfo{year}{2006}).

\bibitem{Tao2011}
\bibinfo{author}{\bibfnamefont{C.}~\bibnamefont{Tao}},
\bibinfo{author}{\bibfnamefont{L.}~\bibnamefont{Jiao}},
\bibinfo{author}{\bibfnamefont{O.~V.}~\bibnamefont{Yazyev}},
\bibinfo{author}{\bibfnamefont{Y.-C.}~\bibnamefont{Chen}},
\bibinfo{author}{\bibfnamefont{J.}~\bibnamefont{Feng}},
\bibinfo{author}{\bibfnamefont{X.}~\bibnamefont{Zhang}},
\bibinfo{author}{\bibfnamefont{R.~B}~\bibnamefont{Capaz}},
\bibinfo{author}{\bibfnamefont{J.~M.}~\bibnamefont{Tour}},
\bibinfo{author}{\bibfnamefont{A.}~\bibnamefont{Zettl}},
\bibinfo{author}{\bibfnamefont{S.~G.}~\bibnamefont{Louie}},
\bibinfo{author}{\bibfnamefont{H.}~\bibnamefont{Dai}}, \bibnamefont{and}
\bibinfo{author}{\bibfnamefont{M.~F}~\bibnamefont{Crommie}},
  \bibinfo{journal}{Nat. Phys.} 1991 (2011).

\bibitem[{\citenamefont{Wimmer et~al.}(2008)\citenamefont{Wimmer, Adagideli,
  Berber, Tom\'anek, and Richter}}]{Wimmer2008}
  \bibinfo{author}{\bibfnamefont{M.}~\bibnamefont{Wimmer}},
  \bibinfo{author}{\bibfnamefont{I.}~\bibnamefont{Adagideli}},
  \bibinfo{author}{\bibfnamefont{S.}~\bibnamefont{Berber}},
  \bibinfo{author}{\bibfnamefont{D.}~\bibnamefont{Tom\'anek}},
  \bibnamefont{and} \bibinfo{author}{\bibfnamefont{K.}~\bibnamefont{Richter}},
  \bibinfo{journal}{Phys. Rev. Lett.} \textbf{\bibinfo{volume}{100}},
  \bibinfo{pages}{177207} (\bibinfo{year}{2008}).

\bibitem[{\citenamefont{Wurm et~al.}(2009{\natexlab{b}})\citenamefont{Wurm,
  Wimmer, Adagideli, Richter, and Baranger}}]{Wurm2009a}
\bibinfo{author}{\bibfnamefont{J.}~\bibnamefont{Wurm}},
  \bibinfo{author}{\bibfnamefont{M.}~\bibnamefont{Wimmer}},
  \bibinfo{author}{\bibfnamefont{I.}~\bibnamefont{Adagideli}},
  \bibinfo{author}{\bibfnamefont{K.}~\bibnamefont{Richter}}, \bibnamefont{and}
  \bibinfo{author}{\bibfnamefont{H.~U.} \bibnamefont{Baranger}},
  \bibinfo{journal}{New J. Phys.} \textbf{\bibinfo{volume}{11}},
  \bibinfo{pages}{095022} (\bibinfo{year}{2009}{\natexlab{b}}).

\bibitem{Bhowmick2010}
\bibinfo{author}{\bibfnamefont{S.}~\bibnamefont{Bhowmick}} \bibnamefont{and} 
\bibinfo{author}{\bibfnamefont{V.~B.}~\bibnamefont{Shenoy}},
  \bibinfo{journal}{Phys. Rev. B} \textbf{\bibinfo{volume}{82}},
  \bibinfo{pages}{155448} (\bibinfo{year}{2010}).

\end{thebibliography}
\end{document}